\documentclass[12pt]{article}
\pdfoutput=1
\usepackage{tikz}
\usetikzlibrary{matrix,arrows,decorations.pathmorphing,decorations.pathreplacing}
\usepackage{jheppub}
\usepackage{amsmath,amssymb,euscript,array,mathrsfs,appendix,ctable,marvosym}
\usepackage{arydshln}
\usepackage{todonotes}
\usepackage{graphicx}
\usepackage[normalem]{ulem}

\input{tcilatex}

\newcommand{\EQ}[1]{\begin{equation}\begin{split} #1
\end{split}\end{equation}}

\title{Lambda Models From Chern-Simons Theories}
\author{David M. Schmidtt\footnote{david@df.ufscar.br}}

\affiliation{Departamento de F\'\i sica, Universidade Federal de S\~ao Carlos, \\
Caixa Postal 676, CEP 13565-905, S\~ao Carlos-SP, Brazil} 

\abstract{In this paper we refine and extend the results of \cite{lambdaCS}, where a connection between the $AdS_{5}\times S^{5}$ superstring lambda model on $S^{1}=\partial D$ and a double Chern-Simons (CS) theory on $D$ based on the Lie superalgebra $\mathfrak{psu}(2,2|4)$ was suggested, after introduction of the spectral parameter $z$. The relation between both theories mimics the well-known CS/WZW symplectic reduction equivalence but is non-chiral in nature. All the statements are now valid in the strong sense, i.e. valid on the whole phase space, making the connection between both theories precise. By constructing a $z$-dependent gauge field in the 2+1 Hamiltonian CS theory it is shown that: i) by performing a symplectic reduction of the CS theory the Maillet algebra satisfied by the extended Lax connection of the lambda model emerges as a boundary current algebra and ii) the Poisson algebra of the supertraces of $z$-dependent Wilson loops in the CS theory obey some sort of spectral parameter generalization of the Goldman bracket. The latter algebra is interpreted as the precursor of the (ambiguous) lambda model monodromy matrix Poisson algebra prior to the symplectic reduction. As a consequence, the problematic non-ultralocality of lambda models is avoided (for any value of the deformation parameter $\lambda \subset [0,1]$), showing how the lambda model classical integrable structure can be understood as a byproduct of the symplectic reduction process of the $z$-dependent CS theory. \\
\begin{flushleft}
\end{flushleft}
}
\begin{document}

\maketitle


\section{Introduction}
Since their discovery \cite{MZ,BPR}, the study of integrable structures in the context of the AdS/CFT correspondence \cite{Large N}
has been one of the most active topics of research in theoretical high energy physics. The duality establishes an equivalence between two apparently different conformal quantum field theories: the ${\cal{N}}=4$ super Yang-Mills (SYM) theory in 4D and the type IIB Green-Schwarz (GS) superstring in the background $AdS_{5}\times S^{5}$ and the presence of integrability on both sides of this holographic duality strongly suggests that it could play a prominent role on an eventual proof of the conjecture, hence its relevance. For a review on the vast topic of the AdS/CFT integrability see \cite{Review} and the references therein.

Usually, a good strategy to better understand a theory is to embed it into a much bigger mathematical body with the objective of having an improved vantage point. In the AdS/CFT situation a possibility is to study separately the deformations of both integrable structures involved in the correspondence  
and then, under the light of the new findings, look for a connection between them in a more systematic manner. In particular, in the gravity side of the duality this corresponds to deforming the GS superstring curved background with the condition that the integrability and the quantum conformal invariance thereof are both preserved at the same time. The last condition means that the background fields must organize themselves into a solution of the type IIB supergravity equations of motion. Both requirements (integrability and conformal invariance) are very rigid constraints and are, in most of the cases, very difficult to satisfy indeed. Besides these two conditions, the $AdS_{5}\times S^{5}$ GS superstring has two major characteristics: i) a first principle quantization scheme is still an unsolved open problem and ii) this theory belongs to the family of the so-called non-ultralocal integrable field theories \cite{Maillet}, meaning that it is outside the reach of the powerful machinery of the quantum inverse scattering method (QISM), also known as the algebraic Bethe ansatz (ABA). 

In this paper we will consider the lambda deformation of the $AdS_{5}\times S^{5}$ GS superstring sigma model but our results will apply, with minor modifications, to other lambda models as well. The lambda deformation was introduced by Sfetsos in \cite{Sfetsos} for the principal chiral model (PCM) and soon after generalized to sigma models on symmetric spaces in \cite{lambda-bos}, to the $AdS_{5}\times S^{5}$ GS supertring (and semi-symmetric spaces) in \cite{lambda-fer} and a couple of years later to the $AdS_{2}\times S^{2}$ hybrid superstring in \cite{hybrid}. In all cases, the deformation preserves the classical integrability of the parent theory and for the important cases of the GS superstring on the backgrounds $AdS_{n}\times S^{n}$ with $n=2,3,5$, their quantum conformal symmetry as well \cite{lambda back,lambda ads3xs3,ads5xs5}, thus providing consistent string theory backgrounds and fulfilling the two important conditions mentioned in the latter paragraph.

Despite their enticing properties, the $AdS_{5}\times S^{5}$ superstring lambda model is still an integrable field theory of the non-ultralocal type (for any value of the deforming parameter $\lambda\subset [0,1]$, a characteristic that is materialized through the Maillet bracket (see \eqref{Maillet})) with a non-ultralocality that persist even after taking the sine-Gordon (SG) limit, i.e. the $\lambda \rightarrow 0$ limit. When applied to the lambda model of the PCM, taking the SG limit is equivalent to implementing the Faddeev-Reshetikhin (FR) ultralocalization mechanism \cite{FR}, where the QISM can be applied successfully to quantize the theory. Then, the $AdS_{5}\times S^{5}$ lambda model (and in general any lambda model on a symmetric and semi-symmetric space) is still outside the grasp of the QISM even in its simplifying $\lambda \rightarrow 0$ limit. The quantization of non-ultralocal 1+1 dimensional integrable field theories has been a longstanding challenging problem and different approaches to handle this situation in diverse scenarios have been considered along the years, see for instance the references \cite{1,2,3,4,5,Alleviating-bos,6,Alleviating-fer,lattice-Poisson,dihedral,QISM-Tim}, but, unfortunately, finding a systematic quantization scheme for treating this kind of theories has been quite elusive.   

In \cite{lambdaCS}, a connection between lambda models and Chern-Simons (CS) theories was suggested based on the well-known existing relation between ordinary (chiral) WZW models and CS theories, where the WZW model turns out to be the CFT living on the boundary of the solid cylinder where the CS theory is defined \cite{Witten-Jones,Seiberg,zoo}. The relation between both theories was worked out mostly on the constrained surface (i.e. in a weak sense), where the lambda model motion takes place and motivated by the possibility of circumventing the non-ultralocality of the lambda models at the cost of increasing the dimensionality of the underlying field theory from 1+1 to 2+1 dimensions and the total number of Hamiltonian constraints by two. All this accompanied by a careful introduction of the spectral parameter of course.

It is the purpose of this paper to deepen and generalize the results of \cite{lambdaCS} and state the connection between both theories in a strong sense, i.e. within the whole phase space and we do this by employing different but complementary approaches (e.g. Hamiltonian and symplectic formulations etc) in order to clarify the results. Indeed, by carefully introducing the spectral parameter and performing a symplectic reduction of the CS theory defined in the 2+1 dimensional solid cylinder to the subset of flat gauge connections on the disc, the resulting 1+1 dimensional field theory living in the boundary (i.e. the cylinder) is the $AdS_{5}\times S^{5}$ integrable lambda model, in the sense that the remaining phase space boundary degrees of freedom obey a current algebra that is actually the Maillet algebra of the lambda model. However, we will also elaborate on how to recover the lambda model action functional from the CS theory as well. In this way, the non-ultralocality of the lambda models is avoided for any value of the deformation parameter $\lambda$. One of the main results of this approach and summarized in the diagram \eqref{alg square} is that the precursor of the would-be Poisson algebra of the lambda model monodromy matrix turns out to be closely related to some sort of spectral parameter extension of the Goldman bracket \cite{Goldman}, which is used to study the intersection properties of homotopy classes of loops on Riemann surfaces (this certainly deserves further study as our theory is by hypothesis defined on the disc having a trivial fundamental group, but see the comments about this issue in the concluding remarks below). Contrary to the exchange algebra of monodromy matrices in non-ultralocal integrable field theories, which is unknown and ambiguous due to their non-ultralocality \cite{Maillet}, the Goldman bracket has been studied for quite a long time and even quantized \cite{Turaev,Nelson1,Nelson2,Nelson3}, mostly within the context of 2+1 dimensional quantum gravity. This opens the possibility for using the vast amount of results available on CS theories to developed a first principle quantization setup (at least) for the $AdS_{5}\times S^{5}$ lambda model. 

The paper is organized as follows. In section \eqref{2.1}, we introduce the lambda model of the $AdS_{5}\times S^{5}$ superstring, display its classical integrable structure and comment on its main properties. In section \eqref{2.2}, we apply the Dirac algorithm to the superstring lambda model in order to prepare the ground to study its integrability in the presence of constraints. In this section we introduce a partial gauge fixing that will facilitate the construction of the extended Lax connection outside the constraint surface which is a subtle situation per se. In section \eqref{2.3}, we construct the extended Lax connection based on two sensible technical conditions: the extended Lax pair should be a strongly flat connection and its associated monodromy matrix must generate first class conserved quantities. During this process a symmetry enhancement occur and the first footprints of the CS gauge theory start to emerge. In section \eqref{2.4}, we comment on the group of dressing transformation and the dressing gauge. This gauge fixes all the first class constraints including the first class parts of the fermionic constraints and has the important job of fixing just the right conjugacy classes of the monodromy matrix corresponding to the local symmetries of the original lambda model. In section \eqref{3.1}, we introduce the Hamiltonian Chern-Simons theory to be considered in the rest of the paper and comment on its properties. In section \eqref{3.2}, we study the CS theory from the symplectic point of view. This approach is particularly useful to clarify the nature of the gauge symmetries in the presence of boundaries and to understand how, under a symplectic reduction, the physical information of the theory is completely contained in its boundary. In section \eqref{3.3}, we study the CS theory from the Hamiltonian theory point of view. This complements the symplectic approach and deals with the other fields entering the definition of the CS action functional. Also, we exploit the extended wave function to find an action functional associated to the boundary degrees of freedom and that is closely related to the original lambda model action functional. Finally, in section \eqref{3.4}, we introduce the spectral parameter $z$ and recover the results of section \eqref{2.4}, showing that the lambda model is the leftover integrable field theory of the symplectic reduction procedure applied to the CS theory. The boundary Kac-Moody currents algebras are recovered at two special values of the spectral parameter, which is a known result for lambda models. Here we introduce the $z$-dependent Wilson loops on the disc and compute their Poisson algebra. The Poisson algebra of traces of Wilson loops when evaluated at the poles of the lambda deformed twisting function, produce a master formula that is behind the usual definitions of the Goldman bracket when specialized to the classical Lie algebras, suggesting a generalization of our approach to the case where the disc is replaced by a genus $g$ Riemann surface $\Sigma_{g}$ with circle boundaries. We finish with some concluding remarks and comment on some directions for future work. The text is as self-contained as possible with the side goal of serving as an introduction to lambda models as well.

\section{Integrable lambda models}
In this section we consider the lambda models from the Hamiltonian theory point of view and study the classical integrability properties thereof. We restrict the discussion to the case of the Green-Schwarz (GS) $AdS_{5}\times S^{5}$ superstring but other lambda models can be studied following exactly the same lines. The Hamiltonian approach and the classical integrable properties of the theory are considered as well.

\subsection{Lambda models} \label{2.1}

Consider the Lie superalgebra $\mathfrak{f=psu(}2,2|4%
\mathfrak{)}$ of $F=PSU(2,2,|4)$ and its $%
\mathbb{Z}
_{4}$ decomposition induced by the automorphism $\Phi $%
\begin{equation}
\Phi (\mathfrak{f}^{(m)})=i^{m}\mathfrak{f}^{(m)},\text{ \ \ }\mathfrak{f=}%
\bigoplus\nolimits_{i=0}^{3}\mathfrak{f}^{(i)},\text{ \ \ }[\mathfrak{f}%
^{(m)},\mathfrak{f}^{(n)}]\subset \mathfrak{f}^{(m+n)\func{mod}4},\text{ \ \ 
}  \label{auto}
\end{equation}%
where $m,n=0,1,2,3$. From this decomposition we define the following twisted loop superagebra%
\begin{equation}
\hat{\mathfrak{f}}=\bigoplus\nolimits_{n\in 
\mathbb{Z}
}\left( \bigoplus\nolimits_{i=0}^{3}\mathfrak{f}^{(i)}\otimes
z^{4n+i}\right) =\bigoplus\nolimits_{n\in 
\mathbb{Z}
}\hat{\mathfrak{f}}^{(n)},  \label{loop superalgebra}
\end{equation}
which is required in order to exhibit the integrable properties of the theory in terms of the spectral parameter $z$. Denote by $G$ the bosonic Lie group associated to $\mathfrak{f}^{(0)}=\mathfrak{su}(2,2)\times \mathfrak{su}(4)$.

The lambda model on the semi-symmetric space $F/G$ is defined by the following action functional\footnote{%
The 1+1 notation used is: $\sigma^{\pm }=\tau\pm \sigma,$ $\partial
_{\pm }=\frac{1}{2}(\partial _{\tau}\pm \partial _{\sigma}),$ $\eta _{\mu \nu
}=diag(1,-1)$, $\epsilon ^{01}=1$, $\delta _{\sigma\sigma ^{\prime }}$=$\delta(\sigma-\sigma^{\prime})$ and $\delta^{\prime} _{\sigma\sigma ^{\prime }}$=$\partial_{\sigma}\delta(\sigma-\sigma^{\prime})$. Also $a_{\pm }=\frac{1}{2}%
(a_{\tau}\pm a_{\sigma})$ and sometimes we use $\tau=\sigma^{0}$ and $\sigma=\sigma^{1}$ interchangeably.} \cite{lambda-fer}
\begin{equation}
S=S_{F/F}(\mathcal{F},A_{\mu })-\frac{k}{\pi }\dint_{\Sigma
}d^{2}\sigma \left\langle A_{+}(\Omega -1)A_{-}\right\rangle ,\text{ \ \ }k\in 
\mathbb{Z}
,  \label{deformed-GS}
\end{equation}%
where $\left\langle \ast ,\ast \right\rangle =STr(\ast ,\ast )$ is the
supertrace in some faithful representation of the Lie superalgebra $\mathfrak{f}$, $\Sigma=S^{1}\times \mathbb{R}$ is the world-sheet manifold parameterized by $(\sigma, \tau)$ and 
$\Omega \equiv \Omega(\lambda)$, where
\begin{equation}
\Omega(z) =P^{(0)}+z P^{(1)}+z ^{-2}P^{(2)}+z^{-1}
P^{(3)}  \label{GS-projector}
\end{equation}
is the omega projector characteristic of the GS superstring. The $P^{(m)}$ are projectors along the graded components $\mathfrak{f}^{(m)}$ of $\mathfrak{f}$.
Above, we have that%
\begin{equation}
S_{F/F}(\mathcal{F},A_{\mu })=S_{WZW}(\mathcal{F})-\frac{k}{\pi }%
\dint_{\Sigma }d^{2}\sigma \left\langle A_{+}\partial _{-}\mathcal{FF}^{-1}-A_{-}%
\mathcal{F}^{-1}\partial _{+}\mathcal{F-}A_{+}\mathcal{F}A_{-}\mathcal{F}%
^{-1}+A_{+}A_{-}\right\rangle ,
\end{equation}%
where $S_{WZW}(\mathcal{F})$ is the usual level $k$ WZW model action\footnote{The prime denotes an extension of the field $\mathcal{F}$ from $\Sigma$ to $B$.}
\begin{equation}
S_{WZW}(\mathcal{F})=-\frac{k}{2\pi }\int\nolimits_{\Sigma }d^{2}\sigma
\left\langle \mathcal{F}^{-1}\partial _{+}\mathcal{FF}^{-1}\partial _{-}%
\mathcal{F}\right\rangle -\frac{k}{4\pi }\int\nolimits_{B}\chi({\mathcal{F}'}) ,\text{ \ \ }%
\chi({\mathcal{F}'}) =\frac{1}{3}\left\langle (\mathcal{F}'^{-1}d\mathcal{F}')%
^{3}\right\rangle .
\end{equation}%

The original GS superstring coupling constant is\footnote{To match with the notation of \cite{lambda background}, take $\kappa^{2}=4\pi g$.} $\kappa^{2}$ and it is related to $k$ through the relation $\lambda ^{-2}=1+\kappa ^{2}/k$. From \eqref{deformed-GS} we realize that the $\lambda$-deformation can be seen as a deformation of the first order formalism or as a deformation of the non-Abelian version of the Buscher approach to T-duality. In the limit $\lambda \rightarrow 1$ with $k\rightarrow \infty$, $\kappa ^{2}$ fixed and $\mathcal{F}=1+\frac{\nu}{k}+...$ expanded around the identity, the first order form of the sigma model is recovered, while for $\lambda \rightarrow 0$ with $k$ fixed and $\kappa ^{2}\rightarrow \infty$, a current-current perturbation of a gauged WZW is produced. In the latter limit the Poisson current algebra develops a Casimir and fixing it to a constant value is equivalent to perform a Pohlmeyer reduction of the sigma model, where the generalized sine-Gordon models emerge. For these reasons, the $\lambda \rightarrow 1$ limit is called the sigma model limit, while the $\lambda \rightarrow 0$ limit is called the sine-Gordon limit.   

The action \eqref{deformed-GS} is invariant under the following gauge and kappa symmetries written collectively as
\begin{equation}
\delta \mathcal{F}=\alpha \mathcal{F}-\mathcal{F}\beta ,\text{ \ \ }\delta
A_{+}=-D_{+}\alpha ,\text{ \ \ }\delta A_{-}=-D_{-}\beta , \label{var fields}
\end{equation}%
where%
\begin{equation}
\alpha =\Omega \epsilon ,\text{ \ \ }\beta =\epsilon ,\text{ \ \ }\epsilon
=\epsilon ^{(0)}+\epsilon ^{(1)}+\epsilon ^{(3)} \label{alpha-beta-kappa}
\end{equation}%
and%
\begin{equation}
\epsilon ^{(1)}=\big[ A_{+}^{(2)},\kappa ^{(1)}\big] _{+},\text{ \ \ }%
\epsilon ^{(3)}=\big[ A_{-}^{(2)},\kappa ^{(3)}\big] _{+}. \label{kappa parameters}
\end{equation}%
Invariance under kappa symmetry requires restoring the 2d world-sheet
metric and finding the appropriate variation of it. The same strategy used in \cite{PR} and applied to the GS sigma model can be employed, with minor modifications, to the
lambda model \eqref{deformed-GS} as well, see \cite{lambda-fer}.

The gauge field equations of motion are given by\footnote{By and abuse of language we refer to $A_{\pm}$ as the gauge field. Which of the components are true gauge fields follows from the Dirac procedure.}%
\begin{equation}
A_{+}=\left( \Omega ^{T}-D^{T}\right) ^{-1}\mathcal{F}^{-1}\partial _{+}%
\mathcal{F},\text{ \ \ }A_{-}=-\left( \Omega -D\right) ^{-1}\partial _{-}%
\mathcal{FF}^{-1},\text{ \ \ }D=Ad_{\mathcal{F}}.  \label{gauge field eom}
\end{equation}%
After putting them back into the action (\ref{deformed-GS}), a
deformation of the non-Abelian T-dual of the GS superstring with respect to the global left action of the supergroup $F$ is
produced. A dilaton is generated in the process but we will not write its explicit form. 

The $\mathcal{F}$ equations of motion, when combined with \eqref{gauge field eom} can be written in two different but equivalent ways
\begin{equation}
\big[ \partial _{+}+\mathscr{L}_{+}(z_{\pm }),\partial _{-}+\mathscr{L}%
_{-}(z_{\pm })\big]=0 , \label{both eom}
\end{equation}
where $z_{\pm}\equiv \lambda^{\pm 1/2}$ and
\begin{equation}
\mathscr{L}%
_{\pm}(z)=I_{\pm}^{(0)}+zI_{\pm}^{(1)}+z^{\pm2}I_{\pm}^{(2)}+z^{-1}I_{\pm}^{(3)}
\label{Light-cone Lax}
\end{equation}%
is a Lax pair satisfying the condition
\begin{equation}
\Phi (\mathscr{L}_{\pm }(z))=\mathscr{L}_{\pm }(iz) \label{automorphism}
\end{equation}
under the action of $\Phi$ in \eqref{auto}.
Then, the lambda model equations of motion follow from the zero curvature condition of $\mathscr{L}_{\pm}(z)$. Explicitly, we have
\begin{equation}
\begin{aligned}
\big[ I_{+}^{(2)},I_{-}^{(1)}\big]  &=&0, \\
D_{-}^{(0)}I_{+}^{(2)}-\big[ I_{+}^{(1)},I_{-}^{(1)}\big]  &=&0, \\
D_{+}^{(0)}I_{-}^{(1)}-D_{-}^{(0)}I_{+}^{(1)}+\big[ I_{+}^{(2)},I_{-}^{(3)}%
\big]  &=&0, \\
\partial _{+}I_{-}^{(0)}-\partial _{-}I_{+}^{(0)}+\big[
I_{+}^{(0)},I_{-}^{(0)}\big] +\big[ I_{+}^{(1)},I_{-}^{(3)}\big] +\big[
I_{+}^{(2)},I_{-}^{(2)}\big] +\big[ I_{+}^{(3)},I_{-}^{(1)}\big]  &=&0,
\\
D_{+}^{(0)}I_{-}^{(3)}-D_{-}^{(0)}I_{+}^{(3)}+\big[ I_{+}^{(1)},I_{-}^{(2)}%
\big]  &=&0, \\
D_{+}^{(0)}I_{-}^{(2)}+\big[ I_{+}^{(3)},I_{-}^{(3)}\big]  &=&0, \\
\big[ I_{+}^{(3)},I_{-}^{(2)} \big]  &=&0,
\end{aligned} \label{eom}
\end{equation}
where $D_{\pm }^{(0)}(\ast )=\partial _{\pm }(\ast )+\big[ I_{\pm
}^{(0)},\ast \big] $. Above, the $I_{\pm}^{(m)}$, are the components of the deformed dual currents defined in terms of the gauge field \eqref{gauge field eom} via
\begin{equation}
I_{+}=\Omega ^{T}(z_{+})A_{+},\text{ \ \ }I_{-}=\Omega^{-1} (z_{-})A_{-},
\end{equation}
or equivalently,
\begin{equation}
I_{\pm }^{(0)}=A_{\pm }^{(0)},\text{ \ \ }I_{\pm }^{(1)}=z_{\mp }A_{\pm
}^{(1)},\text{ \ \ }I_{\pm }^{(2)}=z_{-}^{2}A_{\pm }^{(2)},\text{ \ \ }%
I_{\pm }^{(3)}=z_{\pm }A_{\pm }^{(3)}. \label{dual currents}
\end{equation}

In terms of the Kac-Moody currents defined below in \eqref{KM currents off}, the equations of motion \eqref{gauge field eom} take the form
\begin{equation}
\mathscr{J}_{+}=\frac{k}{2\pi }\big( \Omega ^{T}A_{+}-A_{-}\big) ,\text{
\ \ }\mathscr{J}_{-}=-\frac{k}{2\pi }\big( A_{+}-\Omega A_{-}\big) \label{A eom}
\end{equation}%
and when combined with \eqref{Light-cone Lax} and \eqref{dual currents} imply the relations%
\begin{equation}
\mathscr{L}_{\sigma }(z_{\mp })=\pm \frac{2\pi }{k}\mathscr{J}_{\pm } \label{L and J}
\end{equation}
between the spatial Lax connection and the Kac-Moody currents.
Moreover, the zero curvature condition of the Lax pair is equivalent to the compatibility condition
\begin{equation}
(\partial _{\mu }+\mathscr{L}_{\mu }(z))\Psi (z)=0, \label{compatibility-Lagrangian}
\end{equation}
where $\Psi (z)$ is the so-called wave function. This last equation together with \eqref{A eom} and \eqref{Light-cone Lax} evaluated at the points $z_{\pm}$, allow to express (on-shell) all the Lagrangian fields of the lambda model in terms of the wave function
\begin{equation}
\begin{aligned}
\mathcal{F} &=\Psi (z_{+})\Psi (z_{-})^{-1},\text{ \ \ \ \ \ \
\ \ \ \ }A_{\pm }=-\partial _{\pm }\Psi (z_{\pm })\Psi (z_{\pm })^{-1}, \\
\Omega ^{T}A_{+} &=-\partial _{+}\Psi (z_{-})\Psi (z_{-})^{-1},\text{ \ \ }%
\Omega A_{-}=-\partial _{-}\Psi (z_{+})\Psi (z_{+})^{-1}. \label{solutions-wave}
\end{aligned} 
\end{equation}
By evaluating the action \eqref{deformed-GS} on these solutions to the equations of motion, we obtain the interesting result
\begin{equation}
S_{\text{on-shell}}=S_{WZW}(\Psi (z_{+}))-S_{WZW}(\Psi (z_{-})) \label{action on-shell}
\end{equation}
signaling a phase space decomposition at the special points $z_{\pm}$. This splitting will be explored and exploited heavily in what follows. Actually, the combined kinetic terms cancel each other by virtue of \eqref{solutions-wave} making the contributions to the action $S_{\text{on-shell}}$ purely topological. This behavior will be explained later from the point of view of the Chern-Simons theory. 

Any gauge fixing can be implemented by choosing a specific form of wave function. This was done in \cite{lambda background} for the superstring in the lambda background where the so-called dressing gauge was introduced and used to construct deformations of the known giant magnon solutions. We will come back to this gauge fixing procedure later on.

\subsection{Hamiltonian structure} \label{2.2}

In this subsection we run the Dirac procedure and study the constraint structure of the theory. This step is mandatory in order to construct the extension of the Lax connection outside the constraint surface that ultimately will reflect the integrable properties of the theory. In \cite{lambdaCS}, the spatial component of such an extended Lax connection was introduced without any justification, in this subsection and the next, we will provide the rigorous proof of how both components of the extended connection are obtained.

The phase space associated to the action functional \eqref{deformed-GS} is described by the following phase space coordinates: two currents $\mathscr{J}_{\pm }$ given by
\begin{equation}
\mathscr{J}_{+}=\frac{k}{2\pi }\left( \mathcal{F}^{-1}\partial _{+}\mathcal{%
F+F}^{-1}A_{+}\mathcal{F-}A_{-}\right) ,\text{ \ \ }\mathscr{J}_{-}=-\frac{k}{%
2\pi }\left( \partial _{-}\mathcal{F\mathcal{F}}^{-1}\mathcal{-F}A_{-}%
\mathcal{F}^{-1}\mathcal{+}A_{+}\right)  \label{KM currents off}
\end{equation}%
that obey the relations of two mutually commuting Kac-Moody algebras\footnote{For the Lie (super)-algebra we use the definitions
$\eta _{AB}=\left\langle T_{A},T_{B}\right\rangle ,$ $%
C_{\mathbf{12}}$ = $\eta ^{AB}T_{A}\otimes T_{B}$ and $u_{\mathbf{1}}=u\otimes I,$ $u_{\mathbf{2}}%
=I\otimes u$, etc.}
\begin{equation}
\{\mathscr{J}_{\pm }(\sigma)_{\mathbf{1}},\mathscr{J}_{\pm }(\sigma^{\prime})_{\mathbf{2}}\}=%
-[C_{\mathbf{12}},\mathscr{J}%
_{\pm }(\sigma^{\prime})_{\mathbf{2}}]\delta _{\sigma\sigma^{\prime}}\mp \frac{k}{2\pi }C_{\mathbf{12}}\delta _{\sigma\sigma^{\prime}}^{\prime },%
\text{ \ \ }\{\mathscr{J}_{\pm }(\sigma)_{\mathbf{1}},\mathscr{J}%
_{\mp }(\sigma^{\prime})_{\mathbf{2}}\}=0  \label{super KM algebra}
\end{equation}%
and two conjugated  pairs of fields $(A_{\pm},P_{\mp})$ with Poisson brackets
\begin{equation}
\{P_{\pm }(\sigma)_{\mathbf{1}},A_{\mp }(\sigma^{\prime})_{\mathbf{2}}\}=\frac{1}{2}C_{\mathbf{12}}\delta _{\sigma\sigma^{\prime}}.
\end{equation}
The time flow on this phase space is determined by the canonical Hamiltonian density
\begin{equation}
H_{C}=-\frac{k}{\pi }\big\langle \left( \frac{\pi }{k}\right) ^{2}\left( 
\mathscr{J}_{+}^{2}+\mathscr{J}_{-}^{2}\right) +\frac{2\pi }{k}\left( A_{+}%
\mathscr{J}_{-}+A_{-}\mathscr{J}_{+}\right) +\frac{1}{2}\left(
A_{+}^{2}+A_{-}^{2}\right) -A_{+}\Omega A_{-}\big\rangle \label{canonical H}
\end{equation}%
through the relation 
\begin{equation}
\partial _{\tau }f =\big\{ f ,h_{C}\big\} ,%
\text{ \ \ }h_{C}=\int\nolimits_{S^{1}}d\sigma H_{C}(\sigma
),
\end{equation}
where $f$ is an arbitrary functional of the phase space variables.

Now we consider the Dirac algorithm. There are two primary constraints
\begin{equation}
P_{+}\approx 0,\text{ \ \ }P_{-}\approx 0. 
\end{equation}%
By adding them to the canonical Hamiltonian we construct the total Hamiltonian
\begin{equation}
H_{T}=H_{C}-2\left\langle u_{+}P_{-}+u_{-}P_{+}\right\rangle,
\end{equation}%
where $u_{\pm}$ are arbitrary Lagrange multipliers. 

Stability of the primary constraints under the flow of $H_{T}$ leads to two secondary constraints given by
\begin{equation}
C_{+}=\mathscr{J}_{+}-\frac{k}{2\pi }\left( \Omega ^{T}A_{+}-A_{-}\right)
\approx 0,\text{ \ \ }C_{-}=\mathscr{J}_{-}+\frac{k}{2\pi }\left(A_{+}-\Omega
A_{-}\right) \approx 0,  \label{secondary const}
\end{equation}%
which are nothing but the gauge field equations of motion \eqref{A eom}.
By adding these secondary constraints  to the total Hamiltonian we construct the extended Hamiltonian
\begin{equation}
H_{E}=H_{C}-2\left\langle u_{+}P_{-}+u_{-}P_{+}+\mu_{+}C_{-}+\mu_{-}C_{+}\right\rangle,
\end{equation}
where $\mu_{\pm}$ are arbitrary Lagrange multipliers.

Verifying again the stability of the primary constraints under the flow of $H_{E}$ leads to the conditions
\begin{equation}
\mu_{+}\approx \Omega\mu_{-}\text{ \ \ }\mu_{-}\approx \Omega^{T}\mu_{+}, \label{mus}
\end{equation}
which in turn imply that
\begin{equation}
\mu_{+}^{(0)}\approx \mu_{-}^{(0)},\text{ \ \ }\mu_{+}^{(1)}\approx z_{+}^{2}\mu_{-}^{(1)},\text{ \ \ }\mu_{\pm}^{(2)}\approx 0,\text{ \ \ }\mu_{+}^{(3)}\approx z_{-}^{2}\mu_{-}^{(3)}.
\end{equation}

Now, stability of the secondary constraints under $H_{E}$ gives
\begin{equation}
\begin{aligned}
\partial _{\tau}C_{+} &=\partial _{\sigma}\mathscr{J}_{+}+2\left[ \mathscr{J}%
_{+},A_{-}+\mu _{-}\right] +\frac{k}{\pi }\partial _{\sigma}(A_{-}+\mu _{-})+%
\frac{k}{2\pi }\left( u_{-}-\Omega ^{T}u_{+}\right) \approx 0, \\
\partial _{\tau}C_{-} &=-\partial _{\sigma}\mathscr{J}_{-}+2\left[ \mathscr{J}%
_{-},A_{+}+\mu _{+}\right] -\frac{k}{\pi }\partial _{\sigma}(A_{+}+\mu _{+})+%
\frac{k}{2\pi }\left( u_{+}-\Omega u_{-}\right) \approx 0. \label{sec}
\end{aligned}
\end{equation}%
By using the result \eqref{mus} and by rearranging \eqref{sec} we end up with
\begin{eqnarray}
\left( \Omega \Omega ^{T}-1\right) u_{+} &\approx &\left( \Omega \Omega
^{T}-1\right) \partial _{\sigma}A_{+}+2\Omega \left[ \Omega
^{T}A_{+}-A_{-},A_{-}+\mu _{-}\right]   \\
&&-2\left[ A_{+}-\Omega A_{-},A_{+}+\mu _{+}\right],  \notag \\
\left( \Omega ^{T}\Omega -1\right) u_{-} &\approx &-\left( \Omega ^{T}\Omega
-1\right) \partial _{\sigma}A_{-}-2\Omega ^{T}\left[ A_{+}-\Omega A_{-},A_{+}+\mu
_{+}\right]  \notag \\
&&+2\left[ \Omega ^{T}A_{+}-A_{-},A_{-}+\mu _{-}\right].   \label{grade 2 multipliers}
\end{eqnarray}%
From these expressions we conclude that the Lagrange multipliers $u_{\pm}^{(i)}$, $i=0,1,3$ are completely free, while the $u_{\pm}^{(2)}$ are given in terms of the field content of the theory\footnote{We will not write their explicit form as we will not used them in what follows.}. The projection of these expressions along $\mathfrak{f}^{(0)}$ are trivially satisfied, while the projections along $\mathfrak{f}^{(1)}$ and $\mathfrak{f}^{(3)}$ boils down, for generic values of $\lambda$, to the two following conditions
\begin{equation}
\left[ A_{+}^{(2)},A_{-}^{(1)}+\mu _{-}^{(1)}\right] \approx 0,\text{ \ \ }%
\left[ A_{-}^{(2)},A_{+}^{(3)}+\mu _{+}^{(3)}\right] \approx 0,
\end{equation}%
leading to the solutions%
\begin{equation}
\mu _{-}^{(1)}\approx -A_{-}^{(1)}+\big[ A_{+}^{(2)},\kappa ^{(1)}\big]
_{+},\text{ \ \ }\mu _{+}^{(3)}\approx -A_{+}^{(3)}+\big[
A_{-}^{(2)},\kappa ^{(3)}\big] _{+}. \label{mu sol}
\end{equation}
The terms depending on the arbitrary parameters $\kappa^{(1)}$ and $\kappa^{(3)}$ are associated to the first class parts of the constraints that generate kappa symmetry (cf. \eqref{kappa parameters}). Because of there are no tertiary constraints produced at this level, the algorithm stops here.

Before we consider the Virasoro constraints, it is useful to split the primary and secondary constraints we have found so far into the more relevant separation between first and second class constraints and make some gauge fixings that will simplify the rest of the analysis. 

The first class primary constraints are%
\begin{equation}
P_{+}^{(0)}+P_{-}^{(0)}\approx 0,\text{ \ \ }%
z_{+}P_{+}^{(1)}+z_{-}P_{-}^{(1)}\approx 0,\text{ \ \ }%
z_{-}P_{+}^{(3)}+z_{+}P_{-}^{(3)}\approx 0, \label{1-st class pri}
\end{equation}%
while the second class primary constraints are%
\begin{equation}
P_{+}^{(0)}-P_{-}^{(0)}\approx 0,\text{ \ \ }%
z_{+}P_{+}^{(1)}-z_{-}P_{-}^{(1)}\approx 0,\text{ \ \ }P_{\pm }^{(2)}\approx
0,\text{ \ \ }z_{-}P_{+}^{(3)}-z_{+}P_{-}^{(3)}\approx 0.
\end{equation}%
The second class secondary constraints are given by%
\begin{equation}
C_{-}^{(0)}\approx 0,\text{ \ \ }C_{-}^{(1)}\approx 0,\text{ \ \ }C_{\pm
}^{(2)}\approx 0,\text{ \ \ }C_{-}^{(3)}\approx 0. \label{cons c minus}
\end{equation}

The constraints
\begin{equation}
P_{\pm }^{(2)}\approx 0\text{ \ \ and \ \ }C_{\pm }^{(2)}\approx 0
\end{equation}
form two second class pairs of constraints and we impose them strongly by
means of a Dirac bracket. The brackets between the $\mathscr{J}_{\pm}^{(2)}$ are not modified by virtue of the protection mechanism \cite{lambda-bos}, so we continue using their usual KM Poisson brackets.
Then, we get the strong results
\begin{equation}
\begin{aligned}
I_{+}^{(2)} &=\alpha (z_{-}^{2}\mathscr{J}_{+}^{(2)}+z_{+}^{2}\mathscr{J}%
_{-}^{(2)}),\text{ \ \ }\mathscr{J}_{+}^{(2)}=\frac{k}{2\pi }%
(z_{-}^{2}I_{+}^{(2)}-z_{+}^{2}I_{-}^{(2)}), \\
I_{-}^{(2)} &=\alpha (z_{+}^{2}\mathscr{J}_{+}^{(2)}+z_{-}^{2}\mathscr{J}%
_{-}^{(2)}),\text{ \ \ }\mathscr{J}_{-}^{(2)}=-\frac{k}{2\pi }%
(z_{+}^{2}I_{+}^{(2)}-z_{-}^{2}I_{-}^{(2)}), \label{I two}
\end{aligned}
\end{equation}%
where we have defined\footnote{Not to be confused (in what follows) with the Lie superalgebra element $\alpha$ defined in \eqref{alpha-beta-kappa}.}%
\begin{equation}
\alpha =-\frac{2\pi }{k}\frac{1}{z_{+}^{4}-z_{-}^{4}}.
\end{equation}

In a similar way, the constraints
\begin{equation}
\begin{aligned}
P_{+}^{(0)}-P_{-}^{(0)}&\approx 0\text{ \ \ and \ \ }C_{-}^{(0)}\approx 0,\\
z_{+}P_{+}^{(1)}-z_{-}P_{-}^{(1)}&\approx 0\text{ \ \ and \ \ }C_{-}^{(3)}\approx 0,\\
z_{-}P_{+}^{(3)}-z_{+}P_{-}^{(3)}&\approx 0\text{ \ \ and \ \ }C_{-}^{(1)}\approx 0
\end{aligned}
\end{equation}
are also second class pairs. We set them strongly to zero
and continue using Poisson brackets for the the same reason used right above. Then, we get strongly that
\begin{equation}
I_{1}^{(0)}=-\frac{2\pi }{k}\mathscr{J}_{-}^{(0)},\text{ \ \ }I_{1}^{(1)}=-%
\frac{2\pi }{k}z_{-}\mathscr{J}_{-}^{(1)},\text{ \ \ }I_{1}^{(3)}=-\frac{%
2\pi }{k}z_{+}\mathscr{J}_{-}^{(3)}. \label{I's}
\end{equation}

Now, the first class primary constraints \eqref{1-st class pri} can be gauged fixed by means of the
following gauge fixing conditions\footnote{This in turn imply $A_{+}^{(0)}=A_{0}^{(0)}=A_{1}^{(0)}$, $A_{+}^{(1)}=A_{0}^{(1)}=A_{1}^{(1)}$, $A_{-}^{(3)}=A_{0}^{(3)}=-A_{1}^{(3)}$.
}%
\begin{equation}
A_{-}^{(0)}\approx 0,\text{ \ \ }A_{-}^{(1)}\approx 0,\text{ \ \ }%
A_{+}^{(3)}\approx 0 \label{partial}
\end{equation}%
and we impose them strongly by means of a Dirac bracket, which on phase space functions that are independent of $P_{\pm}$ is equivalent to the Poisson bracket. We will restrict ourselves to this case in what follows.
The last two conditions fix part of the kappa symmetry associated to the
solutions of the Lagrange multipliers\footnote{This partial gauge fixing is natural in the context of the Pohlmeyer reduction of the $AdS_{5}\times S^{5}$ superstring \cite{PR}.} $\mu _{-}^{(1)}$ and $\mu _{+}^{(3)}$
found in \eqref{mu sol}, leaving only the arbitrary parts related to $\kappa _{-}^{(1)}$ and $\kappa _{+}^{(3)}$, see also \eqref{kappa parameters}. Stability of \eqref{partial} under $H_{E}$ also fix some of  the Lagrange multipliers to the values
\begin{equation}
u_{-}^{(0)}\approx 0,\text{ \ \ }u_{-}^{(1)}\approx 0,\text{ \ \ }u_{+}^{(3)}\approx 0.  
\end{equation}

The remaining constraints of the theory are now given by%
\begin{equation}
\begin{aligned}
\varphi ^{(0)}&\equiv C_{+}^{(0)}=\mathscr{J}_{+}^{(0)}+\mathscr{J}_{-}^{(0)}, \\%
\varphi ^{(1)}&\equiv C_{+}^{(1)}=\mathscr{J}_{+}^{(1)}+z_{-}^{2}%
\mathscr{J}_{-}^{(1)}, \\
\varphi ^{(3)}&\equiv C_{+}^{(3)}=\mathscr{J%
}_{+}^{(3)}+z_{+}^{2}\mathscr{J}_{-}^{(3)}. \label{phi const}
\end{aligned}%
\end{equation}
The fermionic constraints $\varphi ^{(1)}$ and $\varphi ^{(3)}$ are
partially first and partially second class as can be seen from
the Poisson algebra of the constraints
\begin{equation}
\begin{aligned}
\big\{\varphi ^{(0)}(\sigma )_{\mathbf{1}},\varphi ^{(0)}(\sigma ^{\prime })_{\mathbf{2}}\big\} 
&=-\big[ C_{\mathbf{12}}^{(00)},\varphi ^{(0)}(\sigma ^{\prime })_{\mathbf{2}}\big] \delta
_{\sigma \sigma ^{\prime }}, \\
\big\{\varphi ^{(0)}(\sigma )_{\mathbf{1}},\varphi ^{(1)}(\sigma ^{\prime })_{\mathbf{2}}\big\} 
&=-\big[ C_{\mathbf{12}}^{(00)},\varphi ^{(1)}(\sigma ^{\prime })_{\mathbf{2}}\big] \delta
_{\sigma \sigma ^{\prime }}, \\
\big\{\varphi ^{(0)}(\sigma )_{\mathbf{1}},\varphi ^{(3)}(\sigma ^{\prime })_{\mathbf{2}}\big\} 
&=-\big[ C_{\mathbf{12}}^{(00)},\varphi ^{(3)}(\sigma ^{\prime })_{\mathbf{2}}\big] \delta
_{\sigma \sigma ^{\prime }}, \\
\big\{\varphi ^{(1)}(\sigma )_{\mathbf{1}},\varphi ^{(1)}(\sigma ^{\prime })_{\mathbf{2}}\big\} 
&=-(z_{-}^{2}/\alpha)\big[ C_{\mathbf{12}}^{(13)},I_{-}^{(2)}(\sigma
^{\prime })_{\mathbf{2}}\big] \delta _{\sigma \sigma ^{\prime }}, \\
\big\{\varphi ^{(1)}(\sigma )_{\mathbf{1}},\varphi ^{(3)}(\sigma ^{\prime })_{\mathbf{2}}\big\} 
&=-\big[ C_{\mathbf{12}}^{(13)},\varphi ^{(0)}(\sigma ^{\prime })_{\mathbf{2}}\big] \delta
_{\sigma \sigma ^{\prime }}, \\
\big\{\varphi ^{(3)}(\sigma )_{\mathbf{1}},\varphi ^{(3)}(\sigma ^{\prime })_{\mathbf{2}}\big\} 
&=-(z_{+}^{2}/\alpha)\big[ C_{\mathbf{12}}^{(31)},I_{+}^{(2)}(\sigma
^{\prime })_{\mathbf{2}}\big] \delta _{\sigma \sigma ^{\prime }}. \label{constraint algebra}
\end{aligned}
\end{equation}
The first class parts being generated by\footnote{To show this it is necessary to impose the Virasoro constraints, c.f. \eqref{Virasoro} and \eqref{constraint surface}.}
\begin{equation}
\varphi _{\bot }^{(1)}=\big[ I_{-}^{(2)},\varphi ^{\left( 1\right) }\big] _{+},\text{ \ \ }\varphi _{\bot }^{(3)}=\big[
I_{+}^{(2)},\varphi ^{\left( 3\right) }\big] _{+}. \label{fer 1-st class}
\end{equation}%

In this partial gauge (i.e. \eqref{partial}), the equations of motion \eqref{eom} and the Lax pair \eqref{Light-cone Lax} take the form
\begin{equation}
\begin{aligned}
\partial _{-}I_{1}^{(0)} &=-\big[ I_{1}^{(1)},I_{1}^{(3)}\big] +\big[
I_{+}^{(2)},I_{-}^{(2)}\big], \\
\partial _{+}I_{1}^{(3)} &=-\big[ I_{1}^{(0)},I_{1}^{(3)}\big] +\big[
I_{1}^{(1)},I_{-}^{(2)}\big], \\
\partial _{-}I_{1}^{(1)} &=-\big[ I_{+}^{(2)},I_{1}^{(3)}\big], \\
\partial _{+}I_{-}^{(2)} &=-\big[ I_{1}^{(0)},I_{-}^{(2)}\big], \\
\partial _{-}I_{+}^{(2)} &=0, \label{partial eom}
\end{aligned}
\end{equation}
and%
\begin{equation}
\mathscr{L}_{+}(z)=I_{1}^{(0)}+zI_{1}^{(1)}+z^{2}I_{+}^{(2)},\text{ \ \ }%
\mathscr{L}_{-}(z)=-z^{-1}I_{1}^{(3)}+z^{-2}I_{-}^{(2)}. \label{Lax gauge-fixed}
\end{equation}
Notice that the components $\mathscr{L}_{\pm}(z)$ are actually expansions around the points $z=0$ and $z=\infty$, respectively. We will exploit this fact later when constructing the extended Lax connection.

Now, we are ready to consider the Virasoro constraints which must be imposed by hand in the conformal gauge approach adopted here. After a temporary reintroduction of the 2d world-sheet metric in the action \eqref{deformed-GS}, we find the stress-tensor components
\begin{equation}
T_{\pm \pm }=-\frac{k}{4\pi }\big\langle \left( \mathcal{F}^{-1}D_{\pm }%
\mathcal{F}\right) ^{2}+2A_{\pm }(\Omega -1)A_{\pm }\big\rangle ,
\end{equation}%
where $D_{\pm }(\ast )=\partial _{\pm }(\ast )+\left[ A_{\pm },\ast \right] .
$ In terms of the constraints $C_{\pm }$ of \eqref{secondary const} they take the quadractic form%
\begin{equation}
\begin{aligned}
T_{++} &=-\frac{k}{4\pi }\big\langle A_{+}(\Omega \Omega
^{T}-1)A_{+}\big\rangle -\big\langle \frac{\pi }{k}%
C_{+}^{2}+C_{+}(\Omega ^{T}-1)A_{+}\big\rangle , \\
T_{--} &=-\frac{k}{4\pi }\big\langle A_{-}(\Omega ^{T}\Omega
-1)A_{-}\big\rangle -\big\langle \frac{\pi }{k}C_{-}^{2}+C_{-}(%
\Omega -1)A_{-}\big\rangle 
\end{aligned}
\end{equation}%
and from this follow another expression for the canonical Hamiltonian \eqref{canonical H}
\begin{equation}
H_{C}=T_{++}+T_{--}-\big\langle A_{0}(C_{+}+C_{-})\big\rangle \label{canonical H 2}
\end{equation}
that will be useful later.

Now, imposing the second class constraints \eqref{cons c minus} strongly in which \eqref{I's} holds and the partial gauge fixing conditions \eqref{partial} imposed so far, we find that
\begin{equation}
\begin{aligned}
T_{++} &=-\frac{1}{2\alpha }\big\langle
I_{+}^{(2)}I_{+}^{(2)}\big\rangle -\big\langle \frac{\pi }{k}\left(
\varphi ^{(0)}\varphi ^{(0)}+2\varphi ^{(1)}\varphi ^{(3)}\right)
-(z_{+}-z_{-})I_{1}^{(1)}\varphi ^{(3)}\big\rangle , \\
T_{--} &=-\frac{1}{2\alpha }\big\langle
I_{-}^{(2)}I_{-}^{(2)}\big\rangle . \label{Virasoro}
\end{aligned}
\end{equation}%
Above, the currents $I_{\pm}^{(2)}$ are given by \eqref{I two}. This last expression for the Virasoro constraints $T_{\pm\pm}\approx0$ is our starting point for the rest of the analysis that follows. 

At this point, the constraints of the theory are given by
\begin{equation}
\varphi ^{(i)}\approx 0\text{ \ \ for \ \ }i=0,1,3\text{ \ \ and \ \ }T_{\pm
\pm }\approx 0 \label{constraint surface}
\end{equation}
with a phase space parameterized only by the Kac-Moody currents $\mathscr{J}_{\pm }$.

It is a well known fact \cite{Kamimura,Das} that because of the fermionic part of the
constraint algebra \eqref{constraint algebra}, the Virasoro constraints $T_{\pm \pm }$ are not first
class and that they must be shifted by terms depending on the
fermionic constraints in order to restore their first class property. Propose
\begin{equation}
T_{++}^{\prime }=T_{++}+\big\langle \lambda ^{(1)}\varphi
^{(3)}\big\rangle, \text{ \ \ } T_{--}^{\prime }=T_{--}+\big\langle
\lambda ^{(3)}\varphi ^{(1)}\big\rangle ,
\end{equation}%
where $\lambda^{(1)}$ and $\lambda^{(3)}$ are to be fixed by requiring the $T'_{\pm\pm}$ to be first class.\
The only problematic Poisson brackets are found to be%
\begin{equation}
\begin{aligned}
\left\{ \varphi ^{(3)}(\sigma ),T_{++}^{\prime }(\sigma ^{\prime })\right\} 
&\approx -(z_{+}^{2}/\alpha )\big[ I_{+}^{(2)},\lambda
^{(1)}+z_{+}I_{1}^{(1)}\big] \delta _{\sigma \sigma ^{\prime }}, \\
\left\{ \varphi ^{(1)}(\sigma ),T_{--}^{\prime }(\sigma ^{\prime })\right\} 
&\approx -(z_{-}^{2}/\alpha )\big[ I_{-}^{(2)},\lambda
^{(3)}-z_{+}I_{1}^{(3)}\big] \delta _{\sigma \sigma ^{\prime }}.
\end{aligned}
\end{equation}%
Then, the solutions are 
\begin{equation}
\lambda ^{(1)}\approx -z_{+}I_{1}^{(1)}+\big[ I_{+}^{(2)},\kappa ^{(1)}\big]
_{+}, \text{ \ \ } \lambda ^{(3)}\approx z_{+}I_{1}^{(3)}+\big[ I_{+}^{(2)},\kappa ^{(3)}\big]
_{+}
\end{equation}%
and include arbitrary terms involving the first class parts of the fermionic
constraints. However, for the sake of constructing an extended Lax pair independent of arbitrary parameters, we will consider instead the following first class Virasoro
constraints given by
\begin{equation}
T_{++}^{\prime }=T_{++}-z_{+}\big\langle I_{1}^{(1)}\varphi
^{(3)}\big\rangle ,\text{ \ \ }T_{--}^{\prime }=T_{--}+z_{+}\big\langle
I_{1}^{(3)}\varphi ^{(1)}\big\rangle .
\end{equation}%
In any case, we can always add the first class part of the fermionic
constraints if needed. Notice that they are automatically stable under the flow of $H_{E}$. They also obey the Virasoro algebra.

We conclude this section by constructing the generator of $\sigma$ translations, i.e. the momentum and by modifying the generator of $\tau$ translations, i.e. the Hamiltonian by terms quadratic in the fermionic constraints. Why we do this modification on the Hamiltonian will be clarified and justified in the next section. 

Because of the phase space is parameterized by the Kac-Moody currents, the momentum generator density must be such that
\begin{equation}
\left\{ \mathscr{J}_{\pm }(\sigma ),P(\sigma ^{\prime })\right\} =\mathscr{J}%
_{\pm }(\sigma^{\prime } )\delta _{\sigma \sigma ^{\prime }}^{\prime }. \label{Momentum}
\end{equation}%
As noticed in \cite{Vicedo-exchange}, in order for this to occur, a term proportional to the bosonic constraint $\varphi^{(0)}$ must be added. We find the momentum density to be given by
\begin{equation}
P=T_{++}^{\prime }-T_{--}^{\prime}-\big\langle I_{1}^{(0)}\varphi
^{(0)}\big\rangle.
\end{equation}
The addition of this term does not spoil the first class property of $P$.

In order to write down the extended momentum and Hamiltonian densities in a canonical Chern-Simons form (we do this below), we must define the following extended stress-tensor components
\begin{equation}
\overline{T}_{++}=T_{++}^{\prime }-\big\langle I_{1}^{(0)}\varphi
^{(0)}\big\rangle+\frac{1}{2}Q,\text{ \ \ }\overline{T}_{--}=T_{--}^{\prime }+\frac{1}{2}Q, \label{extended PH}
\end{equation}
where $Q$ is a term that is at least quadratic in the constraints $\varphi\approx0$ in order to preserve the time flow induced by $H$ and its first class constraint nature. From this we obtain the extended momentum and Hamiltonian
\begin{equation}
\overline{P}=\overline{T}_{++}-\overline{T}_{--}, \text{ \ \ }\overline{H}=\overline{T}_{++}+\overline{T}_{--}. \label{extended H and P}
\end{equation}
The definition of $P$ is not affected by the presence of $Q$ so $\overline{P}=P$. The only relevant modification appears on the definition of $\overline{H}$. However, the only requirement on any extension of the Hamiltonian is that it must reproduce the partially gauge fixed equations of motion \eqref{partial eom} on the constraint surface $\varphi\approx0$. Let us notice that in the absence of the extra term $Q$, the $\overline{H}$ above is nothing but the canonical Hamiltonian \eqref{canonical H 2} restricted to the partially gauge fixed theory so $\overline{H}$ is a very natural quantity (a phase space extension). Its time flow indeed reproduce the equation \eqref{partial eom} when $\varphi\approx0$, as can be seen from the extended equations of motion
\begin{equation}
\begin{aligned}
\partial _{-}I_{1}^{(0)} &=-\big[ I_{1}^{(1)},I_{1}^{(3)}\big] +\big[
I_{+}^{(2)},I_{-}^{(2)}\big] +\alpha z_{+}^{3}\big[ I_{1}^{(1)},\varphi
^{(3)}\big] -\alpha z_{-}^{3}\big[ \varphi ^{(1)},I_{1}^{(3)}\big] , \\
\partial _{+}I_{1}^{(3)} &=-\big[ I_{1}^{(0)},I_{1}^{(3)}\big] +\big[
I_{1}^{(1)},I_{-}^{(2)}\big] +\alpha z_{+}^{3}\partial _{1}\varphi
^{(3)}+\alpha z_{+}^{3}\big[ I_{1}^{(0)},\varphi ^{(3)}\big] +\alpha z_{-}%
\big[ z_{+}^{2}I_{+}^{(2)}-z_{-}^{2}I_{-}^{(2)},\varphi ^{(1)}\big] , \\
\partial _{-}I_{1}^{(1)} &=-\big[ I_{+}^{(2)},I_{1}^{(3)}\big] +\alpha
z_{-}^{3}\partial _{1}\varphi ^{(1)}+\alpha z_{-}^{3}\big[
I_{1}^{(0)},\varphi ^{(1)}\big] +\alpha z_{+}\big[
z_{+}^{2}I_{+}^{(2)}-z_{-}^{2}I_{-}^{(2)},\varphi ^{(3)}\big] , \\
\partial _{+}I_{-}^{(2)} &=-\big[ I_{1}^{(0)},I_{-}^{(2)}\big] -\alpha
z_{+}\big[ I_{1}^{(1)},\varphi ^{(1)}\big] +\alpha z_{+}^{4}\big[
\varphi ^{(0)},I_{-}^{(2)}\big] , \\
\partial _{-}I_{+}^{(2)} &=\alpha \big[ \varphi ^{(0)},I_{-}^{(2)}\big]
+\alpha z_{-}\big[ I_{1}^{(3)},\varphi ^{(3)}\big] , \\
\partial _{-}\varphi ^{(3)} &=-z_{+}\big[ \varphi ^{(0)},I_{1}^{(3)}\big]
+\alpha z_{+}^{4}\big[ \varphi ^{(0)},\varphi ^{(3)}\big] , \\
\partial _{+}\varphi ^{(1)} &=-\big[ I_{1}^{(0)},\varphi ^{(1)}\big]
+\alpha z_{+}^{4}\big[ \varphi ^{(0)},\varphi ^{(1)}\big],  \\ 
\partial _{-}\varphi ^{(0)} &=0,  \label{extended eom}
\end{aligned}
\end{equation}
where we have taken
\begin{equation}
Q=-2\alpha z_{+}^{4} \big\langle \varphi^{(1)}\varphi^{(3)}\big\rangle \label{F}
\end{equation}
in the definition of $\overline{H}$ and organized the $\tau,\sigma$ derivatives in terms of light-cone coordinates derivatives on the lhs.
Why we took this specific form for the term $Q$, will be clarified below. 

Once we have explicit expressions for $\overline{H}$ and $\overline{P}$, we ask if there exist an extension of the Lax pair \eqref{Light-cone Lax} outside the constraint surface $\varphi\approx 0$, such that the equation of motions \eqref{extended eom} follows from its associated zero curvature condition.

\subsection{The extended Lax pair} \label{2.3}

The necessity for introducing the extended Lax pair lies in the fact that the integrability of the theory manifests itself through it rather than through the original Lax pair \eqref{Light-cone Lax} and this is because we are dealing with a constrained integrable field theory as shown by the Dirac procedure above, that is why a thorough analysis of the lambda model phase space structure was necessary. Two criteria are employed now in order to extend the Lax connection outside the constraint surface in a unique way: i) the extended Lax connection should be strongly flat and ii) its associated monodromy matrix should generate conserved integrals of motion that are first class. To accomplish this task we will follow \cite{Vicedo-exchange} quite closely.

In order to construct the extended Lax pair $\overline{\mathscr{L}}_{\pm }(z),
$ we first compute the action of the extended stress tensor components $%
\overline{P}_{\pm }\equiv \overline{T}_{\pm \pm }$ constructed above on the Kac-Moody
currents. We find that such action is given by%
\begin{equation}
\begin{aligned}
\big\{ \mathscr{J}_{+},\int\nolimits_{S^{1}}d\sigma ^{\prime }\overline{P}%
_{\pm }(\sigma ^{\prime })\big\}  &=\frac{k}{2\pi }\partial _{\sigma }%
\overline{\mathscr{L}}_{\pm }(z_{-})+\big[ \mathscr{J}_{+},\overline{%
\mathscr{L}}_{\pm }(z_{-})\big] , \\
\big\{ \mathscr{J}_{-},\int\nolimits_{S^{1}}d\sigma ^{\prime }\overline{P}%
_{\pm }(\sigma ^{\prime })\big\}  &=-\frac{k}{2\pi }\partial _{\sigma }%
\overline{\mathscr{L}}_{\pm }(z_{+})+\big[ \mathscr{J}_{-},\overline{%
\mathscr{L}}_{\pm }(z_{+})\big] , \label{p on J}
\end{aligned}
\end{equation}%
where%
\begin{equation}
\begin{aligned}
\overline{\mathscr{L}}_{+}(z_{-}) &=\mathscr{L}_{+}(z_{-})+(2\pi/k)
\varphi ^{(0)}+\alpha z_{-}^{4}\big( \varphi ^{(1)}+\varphi ^{(3)}\big) ,
\\
\overline{\mathscr{L}}_{-}(z_{-}) &=\mathscr{L}_{-}(z_{-})+\alpha
z_{+}^{4}\big( \varphi ^{(1)}+\varphi ^{(3)}\big) , \\
\overline{\mathscr{L}}_{+}(z_{+}) &=\mathscr{L}_{+}(z_{+})+\alpha \big(
z_{-}^{2}\varphi ^{(1)}+z_{+}^{2}\varphi ^{(3)}\big) , \\
\overline{\mathscr{L}}_{-}(z_{+}) &=\mathscr{L}_{-}(z_{+})+\alpha \big(
z_{-}^{2}\varphi ^{(1)}+z_{+}^{2}\varphi ^{(3)}\big),  \label{relations}
\end{aligned}
\end{equation}%
are extensions of the Lax pair $\mathscr{L}_{\pm }(z)$ given by \eqref{Lax gauge-fixed}
but evaluated at the points $z=z_{\pm }$. Because of the momentum density \eqref{Momentum}
generate $\sigma $-translations, the first conclusion we extract form \eqref{p on J} is that we must still have 
\begin{equation}
\overline{\mathscr{L}}_{\sigma }(z_{\mp })=\pm \frac{2\pi }{k}\mathscr{J}%
_{\pm }. \label{j and l}
\end{equation}
In other words, the relation \eqref{L and J} should be valid outside the constraint surface $\varphi \approx 0$ as well. This result will be quite useful in what follows.

Now, an educated guess for the form of $\overline{\mathscr{L}}_{\sigma }(z)$
obeying the condition \eqref{j and l} and satisfying the property \eqref{automorphism}, in order to be an element of \eqref{loop superalgebra}, is given by \cite{lambdaCS}
\begin{equation}
\overline{\mathscr{L}}_{\sigma }(z)=f_{+}(z)\overline{\Omega }(z/z_{-})%
\mathscr{J}_{+}+f_{-}(z)\overline{\Omega }(z/z_{+})\mathscr{J}_{-}, \label{fundamental}
\end{equation}%
where
\begin{equation}
f_{\pm }(z)=\alpha \left( z^{4}-z_{\pm }^{4}\right) \label{f's}
\end{equation}
and
\begin{equation}
\overline{\Omega }(z)=P^{(0)}+z^{-3}P^{(1)}+z^{-2}P^{(2)}+z^{-1}P^{(3)} \label{Omega Hybrid}
\end{equation}%
is the omega projector characteristic of the lambda model of the hybrid superstring \cite{hybrid}. The Poisson bracket of it with itself is of the Maillet $\mathfrak{r}/\mathfrak{s}$ form \cite{Maillet}
\begin{equation}
\begin{aligned}
\{\overline{\mathscr{L}}_{\sigma}(\sigma;z)_{\mathbf{1}},\overline{\mathscr{L}}_{\sigma}(\sigma^{\prime};w)_{\mathbf{2}}\}=&[\mathfrak{r}
_{\mathbf{12}}(z,w),\overline{\mathscr{L}}_{\sigma}(\sigma;z)_{\mathbf{1}}+\overline{\mathscr{L}}_{\sigma}(\sigma^{\prime};w)_{\mathbf{2}}]\delta
_{\sigma \sigma^{\prime}} \\
+&[\mathfrak{s}_{\mathbf{12}}(z,w),\overline{\mathscr{L}}_{\sigma}(\sigma;z)_{\mathbf{1}}-\overline{\mathscr{L}}_{\sigma}(\sigma^{\prime};w)_{\mathbf{2}}]\delta _{\sigma \sigma^{\prime}}-2\mathfrak{s}_{\mathbf{12}}(z,w)\delta _{\sigma \sigma^{\prime}}^{\prime },
\label{Maillet-lambda}
\end{aligned}
\end{equation}%
where
\begin{equation}
\begin{aligned}
\mathfrak{r}_{\mathbf{12}}(z,w)&=-\frac{1}{z^{4}-w^{4}}\tsum\nolimits_{j=0}^{3}%
\{z^{j}w^{4-j}C_{\mathbf{12}}^{(j,4-j)}\varphi _{\lambda
}^{-1}(w)+z^{4-j}w^{j}C_{\mathbf{12}}^{(4-j,j)}\varphi _{\lambda }^{-1}(z)\}, \\
\mathfrak{s}_{\mathbf{12}}(z,w)&=-\frac{1}{z^{4}-w^{4}}\tsum\nolimits_{j=0}^{3}%
\{z^{j}w^{4-j}C_{\mathbf{12}}^{(j,4-j)}\varphi _{\lambda
}^{-1}(w)-z^{4-j}w^{j}C_{\mathbf{12}}^{(4-j,j)}\varphi _{\lambda }^{-1}(z)\} \label{def r,s}
\end{aligned}
\end{equation}%
and $\varphi _{\lambda }(z)$ is the deformed twisting function given by
\begin{equation}
\varphi _{\lambda }(z)=\frac{2}{\alpha }.\frac{1}{%
(z^{2}-z^{-2})^{2}-(z_{+}^{2}-z_{-}^{2})^{2}}. \label{twisting function}
\end{equation}%
At the points $z=z_{\pm}$, \eqref{Maillet-lambda} reduce to the Kac-Moody algebra we wrote before in \eqref{super KM algebra}. This can be seen from
\begin{equation}
\mathfrak{s}_{\mathbf{12}}(z_{\pm },z_{\pm })=\mp \frac{\pi }{k}C_{\mathbf{12}},\text{ \ \ }%
\mathfrak{s}_{\mathbf{12}}(z_{\pm },z_{\mp })=0 \label{s at poles}
\end{equation}%
and the fact that $\mathfrak{r}_{\mathbf{12}}(z_{\pm },z_{\pm })\sim C_{\mathbf{12}},$ $%
\mathfrak{r}_{\mathbf{12}}(z_{\pm },z_{\mp })=0$. Notice that the points $z=z_{\pm}$ are actually poles of the twisting function. The $\mathfrak{s}_{\mathbf{12}}(z,w)$ above also satisfy 
\begin{equation}
\underset{\lambda \rightarrow 0}{\lim }\ \mathfrak{s}_{\mathbf{12}}(z,w)=-\frac{\pi }{k}%
C_{\mathbf{12}}^{(00)}, \label{SG limit}
\end{equation}
showing that the non-ultralocality persists even in the so-called ultralocalization (sine-Gordon) limit as mentioned before in the introduction.

Inspired by the expansions around $z=0$ and $z=\infty $ of the Lax pair we found before \eqref{Lax gauge-fixed} in the
partial gauge fixing we are considering, we write \eqref{fundamental} in the form 
\begin{equation}
\overline{\mathscr{L}}_{\sigma }(z)=\overline{\mathscr{L}}_{+}(z)-\overline{%
\mathscr{L}}_{-}(z) \label{decomposition}
\end{equation}%
and identify%
\begin{equation}
\begin{aligned}
\overline{\mathscr{L}}_{+}(z) &= f_{+}(z)\mathscr{J}_{+}^{(0)}+
f_{-}(z)\mathscr{J}_{-}^{(0)}+\alpha z\big( z_{-}^{3}\mathscr{J}%
_{+}^{(1)}+z_{+}^{3}\mathscr{J}_{-}^{(1)}\big) \\
& \text{\quad}+\alpha z^{2}\big(z_{-}^{2}\mathscr{J}_{+}^{(2)}+z_{+}^{2}\mathscr{J}_{-}^{(2)}\big)
+\alpha z^{3}\big( z_{-}\mathscr{J}_{+}^{(3)}+z_{+}\mathscr{J}%
_{-}^{(3)}\big)\text{\,}
\end{aligned}
\end{equation}%
and
\begin{equation}
\begin{aligned}
\overline{\mathscr{L}}_{-}(z) &= \alpha z^{-1}\big( z_{+}^{3}\mathscr{J}%
_{+}^{(3)}+z_{-}^{3}\mathscr{J}_{-}^{(3)}\big) +\alpha z^{-2}\big(
z_{+}^{2}\mathscr{J}_{+}^{(2)}+z_{-}^{2}\mathscr{J}_{-}^{(2)}\big) \\
&\text{\quad}+\alpha z^{-3}\big( z_{+}\mathscr{J}_{+}^{(1)}+z_{-}\mathscr{J}_{-}^{(1)}\big) .
\end{aligned}
\end{equation}

Equivalently, in the more familiar light-cone coordinates and in terms of the dual currents, we have
\begin{equation}
\begin{aligned}
\overline{\mathscr{L}}_{+}(z) &=\mathscr{L}_{+}(z)+f_{+}(z)\varphi ^{(0)}+\alpha zz_{-}^{3}\varphi ^{(1)}+\alpha
z^{3}z_{-}\varphi ^{(3)}, \\
\overline{\mathscr{L}}_{-}(z) &=\mathscr{L}_{-}(z)+\alpha
z^{-1}z_{+}^{3}\varphi ^{(3)}+\alpha z^{-3}z_{+}\varphi ^{(1)}. \label{extended Light-cone Lax}
\end{aligned}
\end{equation}%
This extended Lax pair reproduce all four relations of \eqref{relations} when evaluated at the points $z=z_{\pm }$. The strategy has paid off. 

In terms of the dual currents, \eqref{fundamental} takes the form
\begin{equation}
\overline{\mathscr{L}}_{\sigma
}(z)=I_{1}^{(0)}+zI_{1}^{(1)}+z^{2}I_{+}^{(2)}-z^{-2}I_{-}^{(2)}+z^{-1}I_{1}^{(3)}+f_{+}(z)\left( \varphi ^{(0)}+z_{-}^{3}z^{-3}\varphi ^{(1)}+z_{-}z^{-1}\varphi ^{(3)}\right) 
\end{equation}%
and corresponds to the lambda deformation of the analogue quantity used in \cite{Magro-exchange,Vicedo-exchange} to compute the classical exchange algebra for the undeformed $AdS_{5}\times S^{5}$ GS superstring. For the sake of completeness we write the time component as well
\begin{equation}
\overline{\mathscr{L}}_{\tau
}(z)=I_{1}^{(0)}+zI_{1}^{(1)}+z^{2}I_{+}^{(2)}+z^{-2}I_{-}^{(2)}-z^{-1}I_{1}^{(3)}+f_{+}(z)\varphi ^{(0)}+g_{+}(z)\left(z_{-}^{3}z^{-3}\varphi ^{(1)}+z_{-}z^{-1}\varphi ^{(3)}\right),
\end{equation}%
or, equivalently
\begin{equation}
\overline{\mathscr{L}}_{\tau }(z)=g_{+}(z)\overline{\Omega }(z/z_{-})%
\mathscr{J}_{+}+g_{-}(z)\overline{\Omega }(z/z_{+})\mathscr{J}_{-}-2\alpha
\big( z_{+}^{4}\mathscr{J}_{+}^{(0)}+z_{-}^{4}\mathscr{J}_{-}^{(0)}\big), 
\end{equation}
where%
\begin{equation}
g_{\pm }(z)=\alpha \left( z^{4}+z_{\pm }^{4}\right) .
\end{equation}

The components $\overline{\mathscr{L}}_{\mu }(z)$ are not independent and are functions of the same
phase space variables $\mathscr{J}_{\pm }$. This is obvious at this point but will be very important when we consider the Chern-Simons (CS) theory equations of motion, where $\overline{\mathscr{L}}_{\tau }(z_{\pm})$ plays the role of a Lagrange multiplier in that theory. Indeed, the quantity%
\begin{equation}
B(z)=\epsilon ^{\mu \nu }\left\langle \overline{\mathscr{L}}_{\mu }(z)\delta 
\overline{\mathscr{L}}_{\nu }(z)\right\rangle 
\end{equation}%
is independent of $z$ and hence satisfy\footnote{The $z$-dependence of $\mathscr{L}_{\tau }(z)$ is crucial. This will motivate a particular gauge fixing condition in the CS theory.}
\begin{equation}
B(z_{+})-B(z_{-})=0. \label{bdry eom}
\end{equation}%
This condition guarantees that the field configurations obeying the Chern-Simons
equations of motion on the bulk indeed minimize the action functional, i.e. the variational problem is well defined. Recall we have not identified yet the action functional having \eqref{extended eom} as its Euler-Lagrange equations of motion. These extended equations will be identified as the boundary equations of motion in the CS theory to be considered below.

Once we have expressed $\overline{\mathscr{L}}_{\pm }(z)$ in terms of $%
\mathscr{J}_{\pm }$, we can use \eqref{p on J} to compute the difference $\partial _{+}%
\overline{\mathscr{L}}_{-}(z)-\partial _{-}\overline{\mathscr{L}}_{+}(z)$ in
Hamiltonian form. We obtain 
\begin{equation}
\begin{aligned}
\big\{ \overline{\mathscr{L}}_{-}(z),\overline{p}_{+}\big\}-\big\{ 
\overline{\mathscr{L}}_{+}(z),
\overline{p}_{-}\big\}=&\frac{k}{2\pi }f_{-}(z)\overline{\Omega }(z/z_{+})\big[ \overline{%
\mathscr{L}}_{+}(z_{+}),\overline{\mathscr{L}}_{-}(z_{+})\big] \\
&-\frac{k}{2\pi }f_{+}(z)\overline{\Omega }(z/z_{-})\big[ \overline{\mathscr{L}}%
_{+}(z_{-}),\overline{\mathscr{L}}_{-}(z_{-})\big]  \\
=&-\big[ \overline{\mathscr{L}}_{+}(z),\overline{\mathscr{L}}_{-}(z)\big],
\end{aligned}
\end{equation}%
where we have defined
\begin{equation}
\overline{p}_{\pm }=\int\nolimits_{S^{1}}d\sigma \overline{P}_{\pm
}(\sigma ).
\end{equation}
Then, the extended Lax pair $\overline{\mathscr{L}}_{\pm }(z)$ raises as
a strongly flat $z$-dependent connection
\begin{equation}
\partial _{+}\overline{\mathscr{L}}_{-}(z)-\partial _{-}\overline{\mathscr{L}%
}_{+}(z)+\big[ \overline{\mathscr{L}}_{+}(z),\overline{\mathscr{L}}_{-}(z)%
\big] =0 \label{strong flatness}
\end{equation}
and it can be shown that it reproduce the equations of motion we wrote explicitly in \eqref{extended eom}. Notice this result would not be obtained without the presence of the term \eqref{F} in the definition of $\overline{H}$. Equivalently, an extension of the compatibility condition valid now through the whole phase space \eqref{compatibility-Lagrangian} holds
\begin{equation}
(\partial _{\mu }+\overline{\mathscr{L}}_{\mu }(z))\overline{\Psi }(z)=0, \label{ext compa}
\end{equation}%
where $\overline{\Psi }(z)$ is an extension of the wave function $\Psi (z)$. \
Finally, as expected, we get the following result
\begin{equation}
\left\{ \overline{\mathscr{L}}_{\sigma }(\sigma ;z),\overline{T}_{\pm \pm
}(\sigma ^{\prime })\right\} =\overline{\mathscr{L}}_{\pm }(\sigma ^{\prime
};z)\delta _{\sigma \sigma ^{\prime }}^{\prime }-\left[ \overline{\mathscr{L}%
}_{\pm }(\sigma ;z),\overline{\mathscr{L}}_{\sigma }(\sigma ^{\prime };z)%
\right] \delta _{\sigma \sigma ^{\prime }}. \label{flatness alter}
\end{equation}

Generalizing the results of the paper \cite{lambdaCS}, which were valid only on the constraint surface $\varphi \approx 0$, i.e. weakly, we now use \eqref{extended Light-cone Lax} as an input and consider the following Sugawara-type expressions
\begin{equation}
\overline{T}_{\pm \pm }= \pm \frac{k}{4\pi }\big\langle \overline{\mathscr{L}}_{\pm }^{2}(z_{+})-\overline{\mathscr{L}}_{\pm }^{2}(z_{-})\big\rangle . \label{ext stress-tensor}
\end{equation}%
From this definition, we introduce (using \eqref{extended H and P}) the combinations
\EQ{
\overline{H} &=\frac{k}{4\pi }\big\langle \overline{\mathscr{L}}_{\tau }(z_{+})\overline{\mathscr{%
L}}_{\sigma }(z_{+})-\overline{\mathscr{L}}_{\tau }(z_{-})\overline{\mathscr{L}}%
_{\sigma }(z_{-})\big\rangle , \\ 
\overline{P} &=\frac{k}{8\pi }\big\langle (\overline{\mathscr{L}}_{\tau }^{2}(z_{+})+%
\overline{\mathscr{L}}_{\sigma }^{2}(z_{+}))-(\overline{\mathscr{L}}_{\tau }^{2}(z_{-})+\overline{\mathscr{L}}_{\sigma }^{2}(z_{-}))\big\rangle . \label{extended H,P}
}
These two last expressions matches perfectly the extended Hamiltonian and momentum densities constructed in \eqref{extended PH} precisely when $Q$ takes the particular form \eqref{F}, i.e. the addition of $Q$ is also required in order to write the stress tensor $\overline{T}_{\pm\pm}$ entirely as functions of the squares of the components of the extended Lax pair as in \eqref{ext stress-tensor} .

We now turn to the study of the monodromy matrix
\begin{equation}
m(z)=P\exp \big[ -\oint\nolimits_{S^{1}}d\sigma \overline{\mathscr{L}}%
_{\sigma }(\sigma ;z)\big] , \label{monodromy lambda}
\end{equation}%
which, as it is known, is the generator of an infinite number of conserved charges (local and
non-local). At the points $%
z=z_{\pm }$, the associated charges are of the Lie-Poisson type and generate a
Lie-Poisson symmetry that is the classical precursor of a quantum group
symmetry, see \cite{quantum-group} for further details.

Start by calculating the
Poisson bracket of the spatial component of the Lax pair with the constraints, denoted collectively as,%
\begin{equation}
\Phi =( \varphi ^{(i)},T_{\pm \pm }) \text{ \ \ for \ \ }%
i=0,1,3.
\end{equation}%
We find that
\begin{equation}
\begin{aligned}
\big\{\overline{\mathscr{L}}_{\sigma }(\sigma ;z)_{\mathbf{1}},%
\varphi ^{(0)}(\sigma ^{\prime })_{\mathbf{2}}\big\}  &=-C_{\mathbf{12}}^{(00)}\delta _{\sigma
\sigma ^{\prime }}^{\prime }+\big[ C_{\mathbf{12}}^{(00)},\overline{%
\mathscr{L}}_{\sigma }(\sigma ;z)_{\mathbf{1}}\big] \delta _{\sigma \sigma ^{\prime }},
\\
\big\{\overline{\mathscr{L}}_{\sigma }(\sigma ;z)_{\mathbf{1}},%
\varphi ^{(1)}(\sigma ^{\prime })_{\mathbf{2}}\big\} 
&=-z_{-}z^{-1}C_{\mathbf{12}}^{(31)}\delta _{\sigma \sigma ^{\prime }}^{\prime
}+z_{-}z^{-1}\big[ C_{\mathbf{12}}^{(31)},\overline{\mathscr{L}}%
_{\sigma }(\sigma ;z)_{\mathbf{1}}+2z_{+}z\varphi _{\lambda }^{-1}(z)\varphi
^{(1)}(\sigma )_{\mathbf{1}}\big] \delta _{\sigma \sigma ^{\prime }}, \\
\big\{\overline{\mathscr{L}}_{\sigma }(\sigma ;z)_{\mathbf{1}},%
\varphi ^{(3)}(\sigma ^{\prime })_{\mathbf{2}}\big\}  &=-z_{+}zC_{\mathbf{12}}^{(13)}\delta
_{\sigma \sigma ^{\prime }}^{\prime }+z_{+}z\big[ C_{\mathbf{12}}^{(13)},%
\overline{\mathscr{L}}_{\sigma }(\sigma ;z)_{\mathbf{1}}-2\varphi _{\lambda }^{-1}(z)%
\varphi ^{(0)}(\sigma )_{\mathbf{1}}\big] \delta _{\sigma \sigma ^{\prime
}}, \\
\big\{ \overline{\mathscr{L}}_{\sigma }(\sigma ;z),T_{++}(\sigma ^{\prime
})\big\}  &\approx\big(z^{2}I_{+}^{(2)}(\sigma ^{\prime })-z(z_{+}^{2}-1)I_{1}^{(1)}(\sigma ^{\prime })\big)\delta _{\sigma \sigma
^{\prime }}^{\prime } \\
&-\big[ z^{2}I_{+}^{(2)}(\sigma ^{\prime })-z(z_{+}^{2}-1)I_{1}^{(1)}(\sigma ^{\prime }),\overline{%
\mathscr{L}}_{\sigma }(\sigma ;z) \big] \delta
_{\sigma \sigma ^{\prime }}, \\
\big\{ \overline{\mathscr{L}}_{\sigma }(\sigma ;z),T_{--}(\sigma ^{\prime
})\big\}  &\approx z^{-2}I_{-}^{(2)}(\sigma ^{\prime })\delta _{\sigma \sigma
^{\prime }}^{\prime }-z^{-2}\big[ I_{-}^{(2)}(\sigma ^{\prime }),\overline{%
\mathscr{L}}_{\sigma }(\sigma ;z)\big] \delta _{\sigma \sigma ^{\prime }}, \label{L with const}
\end{aligned}
\end{equation}%
where the last two Poisson brackets have been taken weakly and this is because this simplification will not affect a more general discussion to be considered below. The general
form of \eqref{L with const} is then\footnote{By a slight modification, the brackets including $T_{\pm\pm}$ also fit into this form as well. Taking $X_{\mathbf{12}}\rightarrow X_{\pm}$ and $\overline{Y}_{\mathbf{1}}\rightarrow Y_{\pm} $ and dropping all tensor indices.} %
\begin{equation}
\big\{ \overline{\mathscr{L}}_{\sigma }(\sigma ;z)_{\mathbf{1}},%
\Phi (\sigma ^{\prime })_{\mathbf{2}}\big\} =\eta X_{\mathbf{12}}(\sigma ^{\prime })\delta
_{\sigma \sigma ^{\prime }}^{\prime }+\big[ X_{\mathbf{12}}(\sigma ^{\prime }),%
\overline{Y}(\sigma ;z)_{\mathbf{1}}\big] \delta _{\sigma \sigma ^{\prime \label{general PB}
}}.
\end{equation}

The transport matrix is defined by%
\begin{equation}
T( \sigma _{2},\sigma _{1}) =P\exp \big[ -\int\nolimits_{\sigma
_{1}}^{\sigma _{2}}d\sigma \overline{\mathscr{L}}_{\sigma }(\sigma ;z)\big]
\end{equation}%
and its Poisson bracket with the constraints is given by
\begin{equation}
\big\{ T( \sigma _{2},\sigma _{1})_{\mathbf{1}} ,%
\Phi (\sigma ^{\prime })_{\mathbf{2}}\big\} =-\int\nolimits_{\sigma _{1}}^{\sigma
_{2}}d\sigma T( \sigma _{2},\sigma )_{\mathbf{1}} \big\{\overline{\mathscr{L}}_{\sigma }(\sigma ;z)_{\mathbf{1}},\Phi (\sigma
^{\prime })_{\mathbf{2}}\big\} T( \sigma ,\sigma _{1})_{\mathbf{1}} .
\end{equation}%
After making some manipulations using \eqref{general PB} (see \cite{Vicedo-exchange} for the details) and by defining the quantity%
\begin{equation}
\phi _{\epsilon }=\int\nolimits_{S^{1}}d\sigma \big\langle
\epsilon (\sigma )\Phi (\sigma )\big\rangle ,
\end{equation}%
where the gauge parameter $\epsilon$ is independent of the phase space coordinates, we arrive at the following result
\begin{equation}
\left\{ m(z),\phi _{\epsilon }\right\} =\left[ X_{\epsilon },m(z)\right]
-\int\nolimits_{0}^{2\pi }d\sigma ^{\prime }T(2\pi ,\sigma ^{\prime
})\big\langle \big[ X_{\mathbf{12}}(\sigma ^{\prime }),\eta \overline{%
\mathscr{L}}_{\sigma }(\sigma ^{\prime };z)_{\mathbf{1}}+\overline{Y}%
(\sigma ^{\prime };z)_{\mathbf{1}}\big] \epsilon (\sigma ^{\prime
})_{\mathbf{2}}\big\rangle _{\mathbf{2}}T(\sigma ^{\prime },0),
\end{equation}
where we have taken $\sigma_{1}=0$, $\sigma_{2}=2\pi$ and imposed the periodicity condition $\epsilon(\sigma_{2})=\epsilon(\sigma_{1})$. The Lagrangian fields are assumed to be periodic as well. Above, 
\begin{equation}
X_{\epsilon }=-\eta \big\langle X_{\mathbf{12}}(0)\epsilon %
(0)_{\mathbf{2}}\big\rangle _{\mathbf{2}},
\end{equation}%
where $\langle *,* \rangle _{\mathbf{2}}$ instructs us to take the supertrace on the second factor in the tensor product. 

Explicitly, for the constraints $\varphi^{(i)}$, $i=0,1,3$ we obtain, respectively,
\begin{equation}
\begin{aligned}
\left\{ m(z),\phi _{\epsilon ^{(0)}}\right\}  &=\big[ \epsilon
^{(0)}(0),m(z)\big] , \\
\left\{ m(z),\phi _{\epsilon ^{(3)}}\right\}  &=(z_{-}/z)\left[ \epsilon
^{(3)}(0),m(z)\right] -2\varphi _{\lambda }^{-1}(z)\int\nolimits_{0}^{2\pi
}d\sigma ^{\prime }T(2\pi ,\sigma ^{\prime })\left[ \epsilon ^{(3)}(\sigma
^{\prime }),\varphi ^{(1)}(\sigma ^{\prime })\right] T(\sigma ^{\prime },0),
\\
\left\{ m(z),\phi _{\epsilon ^{(1)}}\right\}  &=z_{+}z\left[ \epsilon
^{(1)}(0),m(z)\right] +2z_{+}z\varphi _{\lambda
}^{-1}(z)\int\nolimits_{0}^{2\pi }d\sigma ^{\prime }T(2\pi ,\sigma ^{\prime
})\left[ \epsilon ^{(1)}(\sigma ^{\prime }),\varphi ^{(0)}(\sigma ^{\prime })%
\right] T(\sigma ^{\prime },0), \label{mono gauge}
\end{aligned}
\end{equation}%
while for the $T_{\pm\pm}$, we get
\begin{equation}
\left\{ m(z),\phi _{\epsilon ^{(\pm )}}\right\} \approx \left[ X_{\epsilon
^{(\pm )}},m(z)\right] ,
\end{equation}%
where%
\begin{equation}
X_{\epsilon ^{(\pm )}}=-\eta X_{\pm }(0)\epsilon ^{(\pm )}(0).
\end{equation}%

The expressions \eqref{mono gauge} have a two-fold interpretation. First, when restricted to the constraint surface (at least for $\varphi^{(1)}$ and $\varphi^{(3)}$), the monodromy matrix becomes first class and this means that the surface $\Phi \approx 0$, where the lambda model motion takes place, is preserved by the infinite number of hidden symmetry flows generated by the supertrace of powers of $m(z)$\footnote{This was first observed in \cite{Vicedo-exchange} for the un-deformed $AdS_{5}\times S^{5}$ GS superstring.} . Second, when evaluated at the poles, the three constraints $\varphi^{(i)}$ generate gauge transformations strongly whose finite actions result in conjugations of the monodromy matrix at $z=z_{\pm}$
\begin{equation}
\begin{aligned}
m^{\prime }(z) &=g(0)m(z)g(0)^{-1},\text{ \ \ \ \ \ \ \ \ \ }g=\exp \epsilon ^{(0)},\text{
\ \ for \ \ }\varphi ^{(0)}, \\
m^{\prime }(z_{\pm }) &=g_{\pm }(0)m(z_{\pm })g_{\pm }(0)^{-1},\text{ \ \ }%
g_{+}=\exp z_{-}^{2}\epsilon ^{(3)},\text{ \ \ }g_{-}=\exp \epsilon ^{(3)},%
\text{ \ \ for \ \ }\varphi ^{(1)}, \\
m^{\prime }(z_{\pm }) &=g_{\pm }(0)m(z_{\pm })g_{\pm }(0)^{-1},\text{ \ \ }%
g_{+}=\exp z_{+}^{2}\epsilon ^{(1)},\text{ \ \ }g_{-}=\exp \epsilon ^{(1)},%
\text{ \ \ for \ \ }\varphi ^{(3)}. \label{mini-conjugations}
\end{aligned}
\end{equation}
Despite of the fact that $\varphi^{(1)}$ and $\varphi^{(3)}$ are mixtures of first and second class constraints, their combine action resembles formally a gauge transformation. 
This is a consequence of the Kac-Moody current algebra structure of the phase space and the fact that
\begin{equation}
m(z_{\pm })=P\exp \big[ \pm \frac{2\pi }{k}\int\nolimits_{S^{1}}d\sigma 
\mathscr{J}_{\mp }(\sigma )\big] . \label{monodromy at poles}
\end{equation}%
Then, fixing the gauge symmetries of the theory amounts to fixing these particular conjugacy classes of the monodromy matrix at the poles $z_{\pm}$ with the important proviso that the gauge fixing must fix only the fermionic first class parts of $\varphi^{(1)}$ and $\varphi^{(3)}$ given by \eqref{fer 1-st class}.
The infinitesimal gauge variations are the same as the ones induced by \eqref{var fields} on the currents \eqref{KM currents off} in terms of the Lagrangian fields. By doing this, we give a geometric interpretation to the action of the Hamiltonian constraints on the extended phase space of the lambda model. More on this below.

It is important to realize that as we extend the phase space of the lambda model new gauge symmetries might emerge during the process because we are implicitly modifying the original action functional (the action determine the constraint structure of the theory) through the equations of motion, see \eqref{extended eom} for instance. Indeed, beyond the symmetries coming from the Lagrangian formulation (i.e. the ones generated by the first class constraints in the set $\varphi$), it is evident that the supertraces of powers of \eqref{monodromy at poles} have a larger group of symmetries. To see this, consider the functional
\begin{equation}
\overline{H}(\eta )=-\frac{k}{2\pi }\int_{S^{1}}d\sigma \left\langle \eta _{+}\overline{%
\mathscr{L}}_{\sigma }(z_{+})-\eta _{-}\overline{\mathscr{L}}_{\sigma
}(z_{-})\right\rangle, \label{full gauge gen}
\end{equation}%
where $\eta_{\pm}$ are arbitrary functions. The infinitesimal action given by
\begin{equation}
\begin{aligned}
\{\overline{%
\mathscr{L}}_{\sigma }(z_{\pm }),\overline{H}(\eta )\}&=-\partial _{\sigma }\eta _{\pm
}-[\overline{\mathscr{L}}_{\sigma }(z_{\pm }),\eta _{\pm }], \\
\{m(z_{\pm}),\overline{H}(\eta )\}&=[\eta _{\pm
}(0),m(z_{\pm})]
\end{aligned}
\end{equation}
induce a full action of the supergroup $F=PSU(2,2|4)$
\begin{equation}
\begin{aligned}
\overline{\mathscr{L}}^{\prime }_{\sigma }(z_{\pm })&=g_{\pm }\overline{%
\mathscr{L}}_{\sigma }(z_{\pm })g_{\pm }^{-1}-\partial _{\sigma }g_{\pm
}g_{\pm }^{-1} \\
m^{\prime }(z_{\pm })&=g_{\pm }(0)m(z_{\pm })g_{\pm }(0)^{-1}, \label{full gauge act}
\end{aligned}
\end{equation}%
where $ g_{\pm }=\exp \eta _{\pm }$, that is not predicted by the Dirac procedure considered so far. However,
if we consider the particular choice of gauge parameters
\begin{equation}
\eta _{+}=\Omega \epsilon ,\text{ \ \ }\eta _{-}=\epsilon ,\text{ \ \ }%
\epsilon=\epsilon ^{(0)}+\epsilon ^{(1)}+\epsilon ^{(3)},
\end{equation}%
we do recover \eqref{mini-conjugations} as a subset of gauge transformation. In the next section, we will see how this symmetry enhancement is naturally explained
from the Chern-Simons theory point of view, where the Hamiltonian $\overline{H}$ introduced in \eqref{extended H,P} emerges in a canonical way and the $PSU(2,2|4)$ gauge symmetry predicted by the Dirac procedure. In other words,
the lambda model phase space is embedded into the phase space of a bigger theory which turns out to be of the CS type.

Concerning the Virasoro constraints and keeping in mind the relation with the Chern-Simons theory, we should use $\overline{T}_{\pm\pm}$ instead of the $T_{\pm\pm}$. By using \eqref{flatness alter}, the last two lines in \eqref{L with const} are now replaced by the strong results
\begin{equation}
\left\{ T(\sigma _{2},\sigma _{1}),\overline{T}_{\pm \pm }(\sigma ^{\prime
})\right\} =T(\sigma _{2},\sigma _{1})\overline{\mathscr{L}}_{\pm }(\sigma
^{\prime };z)\delta _{\sigma _{1}\sigma ^{\prime }}-\overline{\mathscr{L}}%
_{\pm }(\sigma ^{\prime };z)T(\sigma _{2},\sigma _{1})\delta _{\sigma
_{2}\sigma ^{\prime }},
\end{equation}
from which follows directly the usual statements concerning the monodromy matrix. In particular, the time conservation of the supertrace of powers of $m(z)$,
\begin{equation}
\left\langle m(z)^{N}\right\rangle ,\text{ \ \ }N\in 
\mathbb{Z}
^{+}
\end{equation}
follows directly from
\begin{equation}
\left\{ T(\sigma _{2},\sigma _{1}),\overline{h}\right\} =T(\sigma
_{2},\sigma _{1})\overline{\mathscr{L}}_{\tau }(\sigma _{1};z)-\overline{%
\mathscr{L}}_{\tau }(\sigma _{2};z)T(\sigma _{2},\sigma _{1}),
\end{equation}
where $\overline{h}=\overline{p}_{+}+\overline{p}_{-}$. Another interesting result is the action of the momentum on the transport
matrix%
\begin{equation}
\left\{ T(\sigma _{2};\sigma _{1}),\overline{p}\right\} =(\partial _{\sigma
_{1}}+\partial _{\sigma _{2}})T(\sigma _{2};\sigma _{1}),
\end{equation}%
where $\overline{p}=\overline{p}_{+}-\overline{p}_{-}$. This last expression is, to our knowledge, not found in the literature.

A comment is in order. We can also apply the direct approach strategy employed in
\cite{Vicedo-exchange} for the construction of the extended Lax pair in which an arbitrary combination of phase space variables is proposed as an input for $\overline{\mathscr{L}}_{\pm}$. By enforcing the conditions that the Lax connection must be strongly flat and that its associated monodromy matrix must be a first class quantity, we arrive to the same answer written in \eqref{extended Light-cone Lax}. The arguments presented here and materialized in the key relations \eqref{fundamental} and \eqref{decomposition} avoids that extra effort. 

\subsection{Dressing group and dressing gauge} \label{2.4}

We now make some observations concerning the group of dressing transformations \cite{dressing} and the dressing gauge that are relevant to the present discussion. For further details and applications to lambda models see \cite{lambda background}. 

In order to understand one of the main properties of the dressing group, let us consider again the action of \eqref{full gauge gen} on the spatial Lax connection. Write \eqref{full gauge gen} in the form
\begin{equation}
\overline{H}(\epsilon )=\int_{S^{1}}d\sigma \left\langle \epsilon \varphi \right\rangle
,\text{ \ \ }\varphi =-\frac{k}{2\pi }\left( \Omega ^{T}\overline{\mathscr{L}%
}_{\sigma }(z_{+})-\overline{\mathscr{L}}_{\sigma }(z_{-})\right) ,
\end{equation}%
where
\begin{equation}
\eta _{+}=\Omega \epsilon ,\text{ \ \ }\eta _{-}=\epsilon ,\text{ \ \ }%
\epsilon =\epsilon ^{(0)}+\epsilon ^{(1)}+\epsilon ^{(2)}+\epsilon ^{(3)}.
\end{equation}%
Its action generalize the first three lines in \eqref{L with const} and \eqref{mono gauge} to%
\begin{equation}
\{\overline{\mathscr{L}}_{\sigma }(\sigma;z),\overline{H}(\epsilon )\}=-\partial _{\sigma
}\epsilon(\sigma;z)-[\overline{\mathscr{L}}_{\sigma }(\sigma;z),\epsilon(\sigma;z)]+2\varphi _{\lambda }^{-1}(z)X(\sigma;z), \label{Pos act}
\end{equation}%
and
\begin{equation}
\{m(z),\overline{H}(\epsilon )\}=[\epsilon (0;z),m(z)]-2\varphi _{\lambda
}^{-1}(z)\int_{0}^{2\pi }d\sigma T(2\pi ,\sigma ;z)X(\sigma ;z)T(\sigma ,0;z),
\end{equation}
respectively. We have defined $\epsilon(\sigma;z)=\Omega (z/z_{-})\epsilon(\sigma)$ and
\begin{equation}
X(\sigma;z)=[z_{+}z\varphi ^{(0)},\epsilon ^{(1)}]-[z_{-}z^{-1}\varphi ^{(1)}+\varphi
^{(2)},\epsilon ^{(2)}]-[\varphi ^{(1)},\epsilon ^{(3)}].
\end{equation}
As noticed before, gauge transformations can not be extended to act (in the strong sense) on $%
\overline{\mathscr{L}}_{\sigma }(z)$ except for those generated by the subalgebra $\mathfrak{f}^{(0)}$ and this is because their action preserve the analytic structure
(i.e. the $z$-dependence) and the character of $\overline{\mathscr{L}}_{\sigma }(z)$ as a gauge connection. The group of dressing transformations extend both properties but to the action of the loop supergroup $\hat{F}$ associated to $F=PSU(2,2|4)$, whose Lie superalgebra $\hat{\mathfrak{f}}$ was written in \eqref{loop superalgebra}. Two consequences of this are: the preservation of the strong flatness condition \eqref{strong flatness} and the preservation of the Virasoro constraints.

Explicitly, a dressing transformation is a map\footnote{%
We use $x$, in this subsection, to denote an arbitrary dependence on the pair of coordinates $(\tau,\sigma)$.}
\begin{equation}
\overline{\Psi }(x;z)\rightarrow \overline{\Psi }^{g}(x;z)=\Theta _{\pm
}(x;z)\overline{\Psi }(x;z)g_{\pm }^{-1}(z), \label{dressing wave}
\end{equation}%
where $\overline{\Psi }(x;z)$ is the wave function in the compatibility
condition \eqref{ext compa}. The elements $\Theta _{\pm }(x;z)$ and $g_{\pm }^{-1}(z)$ are defined
through (well-defined) Riemann-Hilbert factorization problems in the loop supergroup $\widehat{F}$%
\begin{equation}
\Theta (x;z)=\Theta _{-}(x;z)^{-1}\Theta _{+}(x;z),\text{ \ \ }%
g(z)=g_{-}(z)^{-1}g_{+}(z),
\end{equation}%
where%
\begin{equation}
\Theta (x;z)=\overline{\Psi }(x;z)g(z)\overline{\Psi }(x;z)^{-1}.
\end{equation}%
The $\pm $ means projections along the subalgebras $\widehat{\mathfrak{f}}%
_{\geq 0\text{ }}$and $\widehat{\mathfrak{f}}_{<0\text{ }}$of $\widehat{%
\mathfrak{f}}$ formed by elements having integer powers of $z$ that are $\geq 0$ and 
$<0,$ respectively. \

Due to the analytic properties of $\Theta _{\pm },$ we
have that $\Theta _{+}\in \exp \widehat{\mathfrak{f}}_{\geq 0\text{ }}$  and $%
\Theta _{-}\in \exp \widehat{\mathfrak{f}}_{<0\text{ }},$ so we can take%
\begin{equation}
\begin{aligned}
\Theta _{+}(x;z) &=\gamma (x)^{-1}\exp [\sum\limits_{n=1}^{\infty }\theta
_{n}(x)z^{n}], \\
\Theta _{-}(x;z) &=\exp [\sum\limits_{n=1}^{\infty }\theta _{-n}(x)z^{-n}],
\end{aligned}
\end{equation}%
where $\gamma \in G$. \ The dressing group action on the extended Lax pair
is given by%
\begin{equation}
\overline{\mathscr{L}}_{+}^{g}=-\partial _{+}\Theta _{\pm }\Theta _{\pm
}^{-1}+\Theta _{\pm }\overline{\mathscr{L}}_{+}\Theta _{\pm }^{-1},\text{ \
\ }\overline{\mathscr{L}}_{-}^{g}=-\partial _{-}\Theta _{\pm }\Theta _{\pm
}^{-1}+\Theta _{\pm }\overline{\mathscr{L}}_{-}\Theta _{\pm }^{-1},
\end{equation}%
where both $\Theta _{\pm }$ produce the same effect. Notice that the gauge
parameters are field-dependent and non-linear and due to the non-ultralocal
nature of the Maillet bracket \eqref{Maillet-lambda}, the Poisson form of an infinitesimal
dressing transformation is still unknown because the wave function exchange algebra in non-ultralocal integral field theories remains a mistery. In other words, if we take%
\begin{equation}
g(z)=1+\widetilde{X}(z),\text{ \ \ }\widetilde{X}(z)=X_{+}(z)-X_{-}(z)\\
\end{equation}%
and define 
\begin{equation}
\epsilon _{\pm }(x;z)=(\overline{\Psi }(x;z)\widetilde{X}(z)\overline{%
\Psi }(x;z)^{-1})_{\pm },
\end{equation}
we can write the variations as usual (for fixed $\tau$)
\begin{equation}
\begin{aligned}
\delta _{\widetilde{X}}\overline{\Psi} (\sigma;z)&=\epsilon _{\pm }(\sigma;z)\overline{\Psi}(\sigma;z)-\overline{\Psi} (\sigma;z)X_{\pm}(z),\\
\delta _{\widetilde{X}}\overline{\mathscr{L}}_{\sigma }(\sigma;z)&=-\partial _{\sigma
}\epsilon _{\pm }(\sigma;z)-[\overline{\mathscr{L}}_{\sigma }(\sigma;z),\epsilon _{\pm }(\sigma;z)]
\end{aligned}
\end{equation}%
but not in Poisson form as in \eqref{Pos act} above (with $X(\sigma;z)=0$), in the case of the later expression. In any case, the natural gauge action for generic values of $z$ is that of the dressing group, which induce the following change on the monodromy matrix
\begin{equation}
m^{g }(z)=\Theta _{\pm }(2\pi;z)m(z)\Theta _{\pm }(0;z)^{-1}.
\end{equation}%
The dressing factor $\Theta(\sigma;z)$ is not periodic because it depends on the wave function. Hence,
\begin{equation}
m(z)=\overline{\Psi }(2\pi ;z) \overline{\Psi }(0;z)^{-1}\longrightarrow \Theta (2\pi
;z)=m(z)\Theta (0;z)m(z)^{-1}.
\end{equation}%
Yet, $ m^{g}(z)$ leads to conserved charges as well because of $\mathscr{L}^{g}_{\pm} (x;z)$ is a genuine Lax connection for all intents and purposes. Indeed, from \eqref{dressing wave} and the expression right above follows that
\begin{equation}
m^{g}(z)=\overline{\Psi }^{g}(2\pi ;z)\overline{\Psi }^{g}(0;z)^{-1}.
\end{equation}

We now turn to gauge fixing. The dressing gauge fixing condition corresponds to choosing the orbit
of the vacuum solution under the action of the dressing group. Namely,  
\begin{equation}
\overline{\Psi }(x;z)\approx \Theta _{\pm }(x;z)\text{\ }\overline{\Psi }%
_{0}(x;z)g_{\pm }^{-1}(z), \label{dressing gauge}
\end{equation}%
where
\begin{equation}
\overline{\Psi }_{0}(x;z)=\exp
[-(z^{2}\sigma ^{+}+z^{-2}\sigma ^{-})\Lambda ]
\end{equation}%
and the constant element $\Lambda \in \mathfrak{f}^{(2)}$ is such that $\left\langle \Lambda
\Lambda \right\rangle =0$. This choice imply%
\begin{equation}
\begin{aligned}
\overline{\mathscr{L}}_{+}(z) &\approx \gamma ^{-1}\partial _{+}\gamma
+z\psi _{+}+z^{2}\Lambda , \\
\overline{\mathscr{L}}_{-}(z) &\approx z^{-1}\gamma ^{-1}\psi _{-}\gamma
+z^{-2}\gamma ^{-1}\Lambda \gamma ,
\end{aligned}
\end{equation}%
where $\psi _{\pm }\in \func{Im}ad_{\Lambda }.$ When compared to \eqref{extended Light-cone Lax}, we
realize that \eqref{dressing gauge} is actually equivalent to imposing the following collection of gauge fixing conditions 
\begin{equation}
\begin{aligned}
I_{1}^{(0)} &\approx \gamma ^{-1}\partial _{+}\gamma ,
&&I_{1}^{(1)}\approx \psi _{+},\text{ \ \ \ }I_{+}^{(2)}\approx \Lambda , \\
I_{1}^{(3)} &\approx -\gamma ^{-1}\psi _{-}\gamma ,
&&I_{-}^{(2)}\approx \gamma ^{-1}\Lambda \gamma \label{PR}
\end{aligned}
\end{equation}%
supplemented with the vanishing of the constraints
\begin{equation}
\varphi ^{(0)}\approx\varphi ^{(1)}\approx\varphi ^{(3)}\approx 0. \label{PR 2}
\end{equation}%
This is equivalent to perform a Pohlmeyer reduction of the $AdS_{5}\times S^{5}$ GS superstring, see for instance \cite{PR}, for further details. 

The delicate part here is the gauge fixation of the first class constraints \eqref{fer 1-st class}, but fortunately, the fermionic part in \eqref{PR} is also equivalent to
\begin{equation}
\lbrack I_{+}^{(2)},I_{1}^{(1)}]_{+}\approx 0,\text{ \ \ }%
[I_{-}^{(2)},I_{1}^{(3)}]_{+}\approx 0.
\end{equation}%
These are good gauge fixing conditions for the constraints $\varphi _{\perp }^{(3)}\approx 0$ and $\varphi _{\perp
}^{(1)}\approx 0$, respectively, provided that $I_{\pm }^{(2)}I_{\pm
}^{(2)}\approx I$, which is the case. The second class fermionic constraints
are then imposed through \eqref{PR 2}. Concerning the time stability of the dressing gauge \eqref{dressing gauge},
we first notice that it includes the world-sheet coordinates $(\tau ,\sigma )$
explicitly. In these situations, preservation of the gauge fixing condition
is ensured by the emergence of new stress tensor components $\hat{T}%
_{\pm \pm }$ (i.e. new $\hat{H}$ and $\hat{P}$) on the reduced phase space. See \cite{lambda background} for the application to the superstring lambda model and its deformed giant magnons solutions\footnote{Indeed, the lambda-deformed dispersion relation satisfied by the magnon-type solutions is computed in terms of $\hat{H}$ and $\hat{P}$ and an extra $U(1)$ charge $Q$, see \cite{lambda background}.} and \cite{Evans-Tuckey} for the original discussion and formulation from the point of view of symplectic geometry. 

In order to impose \eqref{PR} and \eqref{PR 2} strongly we must introduce a Dirac bracket.
However, for generic values of the deformation parameter $\lambda \in
\lbrack 0,1]$ the bracket is non-local \cite{lambda-bos} and hence not very useful for explicit computations. A local (and 2d relativistic)
Poisson bracket on the reduced phase space is obtained in the $\lambda
\rightarrow 0$ limit and corresponds to the Poisson structure of a fermionic
extension of the non-Abelian Toda model, or generalized sine-Gordon theory,
that appear in the Pohlmeyer reduction of the GS superstring \cite{PR,susy-flows,semi-sym-sG}. This is why taking $\lambda \rightarrow 0$ is called the sine-Gordon limit. The subset of conjugacy classes \eqref{mini-conjugations} of the full $PSU(2,2|4)$ supergroup are, finally, completely gauge fixed by the dressing gauge. 

With all these suggestive results at hand, we now consider the CS theory behind the phase space of the extended  superstring lambda model.

\section{Chern-Simons theories}

In this section we introduce the action functional that reproduce \eqref{strong flatness} as its Euler-Lagrange equation of motion evaluated at the poles $z=z_{\pm}$. This is shown by fixing a particular form of the Lagrange multiplier $A_{\tau}$ in a Hamiltonian Chern-Simons theory on $D \times \mathbb{R}$ to be introduced below and by solving explicitly the Hamiltonian constraints for the curvature of the gauge field $A$ on the disc $D$. The symplectic and Hamiltonian approaches as well as the introduction of the spectral parameter are considered in this section.  

\subsection{Double Chern-Simons action} \label{3.1}

Consider the following double Chern-Simons action functional defined by
\begin{equation}
S_{CS}=S_{(+)}+S_{(-)}, \label{double action}
\end{equation}
where 
\begin{equation}
S_{(\pm )}=\pm \frac{k}{4\pi }\int\nolimits_{M}\big\langle B_{(\pm )}\wedge \hat{d} B_{(\pm )}+\frac{2}{%
3}B_{(\pm )}\wedge B_{(\pm )}\wedge B_{(\pm
)}\big\rangle .  \label{CS copies}
\end{equation}%
The $(\pm )$ sub-index is just a label whose significance will emerge later on, $M
$ is a 3-dimensional manifold and $B_{(\pm )}$ are two different
3-dimensional gauge fields valued in the Lie superalgebra $\mathfrak{f}$. In what follows, we will study
the generic action    
\begin{equation}
S=\frac{\overline{k}}{4\pi }\int\nolimits_{M}\big\langle B\wedge%
\hat{d} B+\frac{2}{3}B\wedge B \wedge B\big\rangle ,%
\text{ \ \ }\overline{k}=\pm k\text{ \ \ for \ \ }(\pm )
\end{equation}
thus to avoid a duplicated analysis.

Under gauge transformations%
\begin{equation}
B^{\prime }=gBg^{-1}-\hat{d}gg^{-1},\text{ \ \ }g\in F,
\end{equation}%
the action changes as follows%
\begin{equation}
S^{\prime }-S=\frac{\overline{k}}{4\pi }\int\nolimits_{M}\chi (g)-\frac{%
\overline{k}}{4\pi }\int\nolimits_{\partial M}\big\langle Bg^{-1}\hat{d}%
g\big\rangle ,
\end{equation}%
where%
\begin{equation}
\chi (g)=\frac{1}{3}\big\langle g^{-1}\hat{d}g\wedge g^{-1}\hat{d}%
g\wedge g^{-1}\hat{d}g\big\rangle .
\end{equation}%
Then, the theory is gauge invariant\footnote{In the sense that $e^{i(S^{\prime}-S)}=1$.} provided%
\begin{equation}
\frac{\overline{k}}{4\pi }\int\nolimits_{M}\chi (g)=2\pi N\text{ \ \ and \ \ 
}g|_{\partial M}=Id, \label{gauge conditions}
\end{equation}%
where $N\in 
\mathbb{Z}
$. We will recover the last condition for gauge invariance from the symplectic approach below.

In order to define the Hamiltonian theory of our interest we consider the
action on the manifold $M=D\times 
\mathbb{R}
$, where $D$ is a 2-dimensional disc parameterized by $x^{i},$ $i=1,2$ and $%
\mathbb{R}
$ is the time direction parameterized by $\tau $. It is useful to use
radius-angle (polar) coordinates $(r,\sigma )$ to describe $D$ as well$.$ In
particular, $\sigma $ is the coordinate along $\partial D=S^{1}$ that is identified with the $S^{1}$ in the definition of lambda model world-sheet $\Sigma=S^{1}\times \mathbb{R}$ in \eqref{deformed-GS}. i.e. the closed string world-sheet corresponds to the boundary of the solid cylinder where the CS theory is defined.

Using the decomposition%
\begin{equation}
B=d\tau A_{\tau }+A,\text{ \ \ }\hat{d}=d\tau \partial _{\tau
}+d ,
\end{equation}%
we end up with the following action functional
\begin{equation}
S=\frac{\overline{k}}{4\pi }\int\nolimits_{D\times 
\mathbb{R}
}d\tau \left\langle -A\partial _{\tau }A+2A_{\tau }F\right\rangle -\frac{%
\overline{k}}{4\pi }\int\nolimits_{\partial D\times 
\mathbb{R}
}d\tau \left\langle A_{\tau }A\right\rangle , \label{action on DxR}
\end{equation}%
where $F=dA+A^{2}$ is the curvature of the 2-dimensional gauge field $A=A_{i}dx^{i}$ not to be confused with the world-sheet gauge field entering the definition of the action \eqref{deformed-GS}.
We also omit the wedge product symbol $\wedge $ in order to simplify
the notation. It is also useful to work in terms of differential forms rather than in terms of components.

The Lagrangian is given by%
\begin{equation}
L=\frac{\overline{k}}{4\pi }\int\nolimits_{D}\left\langle -A\partial _{\tau
}A+2A_{\tau }F\right\rangle -\frac{\overline{k}}{4\pi }\int\nolimits_{%
\partial D}\left\langle A_{\tau }A\right\rangle , \label{double CS Lag}
\end{equation}%
whose arbitrary variation is as follows
\begin{equation}
\delta L=\frac{\overline{k}}{2\pi }\int\nolimits_{D}\left\langle \delta
A_{\tau }F+\delta A(DA_{\tau }-\partial _{\tau }A)\right\rangle +\frac{%
\overline{k}}{4\pi }\int\nolimits_{\partial D}d\sigma \left\langle \delta
A_{\sigma }A_{\tau }-\delta A_{\tau }A_{\sigma }\right\rangle ,
\end{equation}%
where $D(\ast )=d(\ast )+[A,\ast ]$ is a covariant derivative. From this
we find the bulk equations of motion%
\begin{equation}
F=0,\text{ \ \ }\partial _{\tau }A=DA_{\tau },\text{ \ \ on \ \ }D
\label{bulk EOM}
\end{equation}%
stating that the 3-dimensional gauge field $B$ is flat, as well as
the boundary equations of motion 
\begin{equation}
\left\langle \delta A_{\sigma }A_{\tau }-\delta A_{\tau }A_{\sigma
}\right\rangle =0\text{ \ \ on \ \ }\partial D,  \label{Boundary EOM}
\end{equation}%
which must be solved consistently in order to obtain the field configurations minimizing the action. However, as anticipated before, for the lambda models they are automatically satisfied. See, for instance, the discussion around \eqref{bdry eom}.

\subsection{Symplectic approach} \label{3.2}

The symplectic approach to CS theory is particularly useful for understanding the phase space structure and the symmetries in a clear geometrical way. In this section we follow closely the references \cite{Atiyah-Bott,Audin}, where the results of \cite{Audin} valid on genus $g$ Riemann surfaces $\Sigma_{g}$ with boundaries are simplified to the disc (a single boundary). Set $\mathfrak{f}\rightarrow \mathfrak{g}$ throughout this section. A comment is in order, we assume the results are valid for the superalgebra $\mathfrak{psu}(2,2|4)$, so caution is advised as we will ignore any of the subtleties proper to CS theories defined on Lie supergroups in what follows.

The symplectic form associated to the phase space $\mathcal{A}$ of the Lagrangian \eqref{double CS Lag} is given by the Atiyah-Bott 2-form\footnote{Actually, to show this it is necessary to introduce a Dirac bracket first, see the next subsection.}
\begin{equation}
\omega =\frac{\overline{k}}{4\pi }\int\nolimits_{D}\big\langle \delta
A\wedge \delta A\big\rangle \text{ \ \ , \ \ } A\in \mathcal{A}. \label{symplectic CS}
\end{equation}%
The action of the gauge group $\mathcal{G}$ preserves this symplectic form and is
Hamiltonian. To find its corresponding moment map we recall the action of gauge transformations%
\begin{equation}
g\cdot A=gAg^{-1}-dgg^{-1} \label{gauge}
\end{equation}%
and take $g=e^{\eta }$, where $\eta \in \mathbf{g}=\Omega ^{(0)}(D,\mathfrak{g})$. Then, we have%
\begin{equation}
\delta _{\eta }A=-D\eta \longrightarrow X_{\eta }=-(D_{i}\eta )^{A}\frac{\delta }{%
\delta A_{i}^{A}}, \label{Ham vec field}
\end{equation}%
where  $X_{\eta }$ is the associated Hamiltonian vector field. Using the contraction%
\begin{equation}
\delta A(X_{\eta })=-D\eta 
\end{equation}%
in \eqref{symplectic CS} we obtain the desired result%
\begin{equation}
i_{X_{\eta }}\omega =\delta H(\eta ) \text{ \ \ , \ \ } H(\eta )\in \mathcal{C}^{\infty }(\mathcal{A}),
\end{equation}%
where%
\begin{equation}
H(\eta )=\frac{\overline{k}}{2\pi }\int\nolimits_{D}\left\langle \eta
F\right\rangle -\frac{\overline{k}}{2\pi }\int\nolimits_{\partial
D}\left\langle \eta A\right\rangle \label{gauge Hamiltonian}
\end{equation}%
is the gauge Hamiltonian. A second contraction reveals that their Poisson algebra%
\begin{equation}
\{H(\eta ),H(\overline{\eta })\}=H([\eta ,\overline{\eta }])+\frac{\overline{%
k}}{2\pi }\int\nolimits_{\partial D}\left\langle \eta d\overline{\eta }%
\right\rangle  \label{gauge Ham PB}
\end{equation}%
has an extra boundary contribution and this means that the mapping $\eta \rightarrow H(\eta)$ is not a morphism of Lie algebras. However, this can be fixed by equipping $\widehat{\mathbf{g}}=\mathbf{g\oplus 
\mathbb{C}
}$ with the cocycle%
\begin{equation}
c(\eta ,\overline{\eta })=\frac{\overline{k}}{2\pi }\int\nolimits_{\partial
D}\left\langle \eta d\overline{\eta }\right\rangle \label{CS cocycle}
\end{equation}%
and the bracket%
\begin{equation}
\left[ (\eta ,t),(\overline{\eta },s)\right] =([\eta ,\overline{\eta }%
],c(\eta ,\overline{\eta })) .
\end{equation}%
Thus, $\widehat{\mathbf{g}}$ becomes a Lie algebra central extension of $\mathbf{g}$. With the definition
\begin{equation}
H(\eta ,t)=H(\eta )+t, \label{H+t}
\end{equation} 
the mapping%
\begin{equation}
\begin{aligned}
\widehat{\mathbf{g}} &\longrightarrow &\mathcal{C}^{\infty }(\mathcal{A}) \\
(\eta ,t) &\longmapsto &H(\eta ,t)
\end{aligned}%
\end{equation}
becomes a morphism of Lie algebras but with $\widehat{\mathbf{g}}$ acting
infinitesimally on $\mathcal{A}$ in the same way as described above \eqref{Ham vec field}. In particular, the
Hamiltonian vector fields are the same.

This allows to define a non-degenerate pairing between $\Omega ^{2}(D,%
\mathfrak{g})\oplus \Omega ^{1}(\partial D,\mathfrak{g})\oplus 
\mathbb{C}
$ and $\widehat{\mathbf{g}}$ via, e.g. a relation of the type%
\begin{equation}
\left\langle(F,A,z),(\eta ,t)\right\rangle\longrightarrow \frac{\overline{k}}{2\pi }%
\int\nolimits_{D}\left\langle \eta F\right\rangle -\frac{\overline{k}}{2\pi }%
\int\nolimits_{\partial D}\left\langle \eta A\right\rangle +zt. \label{pairing}
\end{equation}%
In this way we identify $\Omega ^{2}(D,\mathfrak{g})\oplus \Omega
^{1}(\partial D,\mathfrak{g})\oplus 
\mathbb{C}
$ as a subspace of $\widehat{\mathbf{g}}^{\ast }.$ Then, the mapping%
\begin{equation} 
\begin{split}
\mu\ :\quad &\mathcal{A} \longrightarrow \ \widehat{\mathbf{g}}^{\ast } \\
&A \longmapsto \ (F,A|_{\partial D},1)
\end{split}
\end{equation}
is an equivariant moment map for the gauge group action. Indeed, under \eqref{gauge} we see from \eqref{H+t} that
\begin{equation}
H(g\cdot A|\eta ,t)=H(A|g^{-1}\eta g,g^{-1}\cdot t), \label{equivariance}
\end{equation}%
where%
\begin{equation}
g^{-1}\cdot t=t+\frac{\overline{k}}{2\pi }\int\nolimits_{\partial D}\left\langle
\eta dgg^{-1}\right\rangle .
\end{equation}

The following normal subgroup of the gauge group $\mathcal{G}$ defined by%
\begin{equation}
\mathcal{G}_{0}=\{g\in \mathcal{G}\text{ }|\text{ }g|_{\partial D}=Id\} \label{normal subgroup}
\end{equation}%
is special (c.f. \eqref{gauge conditions}). It acts on $\mathcal{A}$ with the moment map $A\rightarrow F$
and the space $\mathcal{M}_{0}$ of flat connections modulo the $\mathcal{G}%
_{0}$-action is symplectic. To see this, consider the Lie algebra of $%
\mathcal{G}_{0}$%
\begin{equation}
\mathbf{g}_{0}=\{\eta \in \mathbf{g}\text{ }|\text{ }\eta |_{\partial D}=0\}
\end{equation}%
and embed%
\begin{equation}
\Omega ^{2}(D,\mathfrak{g})\oplus 
\mathbb{C}
\hookrightarrow \widehat{\mathbf{g}}_{0}^{\ast }.
\end{equation}%
From this follows that the moment map for the $\mathcal{G}_{0}$-action is
the map composition 
\begin{equation}
\begin{split}
&\mathcal{A}\ \overset{\mu }{\mathcal{\longrightarrow }}\qquad \ \ \widehat{\mathbf{g}}%
^{\ast }\qquad \ \ \overset{p}{\mathcal{\longrightarrow }}\quad\ \widehat{\mathbf{g}}_{0}^{\ast } \\
&A\ \longmapsto\ (F,A|_{\partial D},1)\ \longmapsto\ (F,1).
\end{split}
\end{equation}%
Hence $\mathcal{M}_{0}=(p\circ \mu )^{-1}(0,1)/\mathcal{G}_{0}$ is a
symplectic reduced space. 

Because of the group $\mathcal{G}_{0}$ is normal in $\mathcal{G}$, the gauge
group also acts on $\mathcal{M}_{0},$ that action preserves the symplectic
structure and the quotient $\mathcal{M}_{0}/\mathcal{G}$ is the CS theory
moduli space $\mathcal{M}$. The latter inherits a natural Poisson structure
coming from the space $\mathcal{A}$. We now proceed to identify this
Poisson structure and its symplectic foliation. Consider the action of the
group $H=\mathcal{G}/\mathcal{G}_{0}$ on $\mathcal{M}_{0}$ and its moment map%
\begin{equation}
\mu \text{ }:\text{ }\mathcal{M\longrightarrow }\ \mathfrak{h}^{\ast }.
\end{equation}%
The poisson bracket on $\mathcal{M}_{0}$ defines a Poisson bracket on $%
\mathcal{M}_{0}/H$ whose symplectic leaves are in one-to-one correspondence
(via $\mu $) with the co-adjoint orbits of $H$ in the image $\mu (\mathcal{M}%
_{0})\subset \mathfrak{h}^{\ast }.$ To see this, consider and element $\xi
\in \mathfrak{h}^{\ast }$ and its orbit $\mathcal{O}_{\xi }.$ Suppose that $H
$ acts on $\mathcal{M}_{0}\times \mathcal{O}_{\xi }$ in a way that%
\begin{equation}
\begin{split}
J :\ \mathcal{M}_{0}\times \mathcal{O}_{\xi }&\longrightarrow \mathfrak{h}^{\ast } \\
\  (A,\alpha )\  &\longmapsto \mu (A)-\alpha 
\end{split}
\end{equation}%
is the moment map for this action. We now perform the symplectic reduction
of $\mathcal{M}_{0}\times \mathcal{O}_{\xi }$ for $J=0.$ Then,%
\begin{equation}
J^{-1}(0)/H=\{(A,\alpha )\in \mathcal{M}_{0}\times \mathcal{O}_{\xi }\text{ }%
|\text{ }\mu (A)=\alpha \}/H\longrightarrow \mu ^{-1}(\mathcal{O}_{\xi })/H \label{isomorphism}
\end{equation}%
is an isomorphism and this defines a symplectic structure on the latter
space. 

We now materialize this discussion.
The symplectic form on $\mathcal{M}_{0}$, denoted by $\omega_{0}$, is computed by the restriction or pull-back of the symplectic form \eqref{symplectic CS} to the subspace of flat connections under the map $p\circ \mu$. Namely,
\begin{equation}
\omega _{0}=\omega |_{A=-d\Psi \Psi ^{-1}},
\end{equation}%
or equivalenty,
\begin{equation}
\omega _{0}=-\frac{\overline{k}}{4\pi }\int\nolimits_{\partial D}d\sigma
\left\langle \delta A_{\sigma }\wedge D_{\sigma }^{-1}(\delta A_{\sigma
})\right\rangle .
\end{equation}%
It is defined only on $\partial D$ and is preserved by the gauge transformations in $H$. Similarly as above, the Hamiltonian vector fields are given by%
\begin{equation}
\delta _{\eta }A_{\sigma }=-D_{\sigma }\eta\longrightarrow X_{\eta }=-(D_{\sigma
}\eta)^{A}\frac{\delta }{\delta A_{\sigma }^{A}}
\end{equation}%
and from this follows that%
\begin{equation}
i_{X_{\eta }}\omega _{0}=\delta H(\eta ),
\end{equation}%
where%
\begin{equation}
H(\eta )=-\frac{\overline{k}}{2\pi }\int\nolimits_{\partial D}
\left\langle \eta A\right\rangle  \label{bdry gauge Ham}
\end{equation}%
is the (boundary) gauge Hamiltonian. A second contraction shows their Poisson algebra%
\begin{equation}
\begin{aligned}
\{H(\eta ),H(\overline{\eta })\} &=\frac{\overline{k}}{2\pi }%
\int\nolimits_{\partial D}d\sigma \big\langle \eta D_{\sigma }\overline{%
\eta }\big\rangle  \\
&=\frac{\overline{k}}{2\pi }\int\nolimits_{\partial D}d\sigma d\sigma
^{\prime }\big\langle \big( C_{\mathbf{12}}\delta _{\sigma \sigma ^{\prime
}}^{\prime }+\big[ C_{\mathbf{12}},A_{\sigma }(\sigma ^{\prime
})_{\mathbf{2}}\big] \delta _{\sigma
\sigma ^{\prime }}\big) ,\eta_{\mathbf{1}} \overline{\eta }_{\mathbf{2}}\big\rangle
_{\mathbf{12}}.
\end{aligned}
\end{equation}%
Then, the theory on $\mathcal{M}_{0}$ is described by a field theory on the boundary $\partial D$ that is endowed with a Poisson structure that is actually a Kac-Moody algebra
\begin{equation}
\{A_{\sigma }(\sigma )_{\mathbf{1}},A_{\sigma }(\sigma ^{\prime
})_{\mathbf{2}}\}=\frac{2\pi}{\overline{k}}(\big[ C_{\mathbf{12}},A_{\sigma }(\sigma ^{\prime
})_{\mathbf{2}}\big] \delta _{\sigma \sigma
^{\prime }}+C_{\mathbf{12}}\delta _{\sigma \sigma ^{\prime }}^{\prime }) \label{CS KM}
\end{equation}
and this is because the components of $A_{\sigma}$ are interpreted as the phase space coordinates of the theory living on $\partial D$. 

From \eqref{bdry gauge Ham}, we can identify $\Omega ^{1}(\partial D,\mathfrak{g})$ with a subspace of $%
\mathfrak{h}^{\ast }$ via, e.g. the pairing%
\begin{equation}
\left\langle A, \eta \right\rangle \longrightarrow -\frac{\overline{k}}{2\pi }\int_{\partial
D}\left\langle \eta A\right\rangle .
\end{equation}%
In this way the equivariant moment map for the action of $\mathcal{G}/\mathcal{G}_{0}$
on $\mathcal{M}_{0}$ is given by
\begin{equation}
\begin{split}
\mathcal{M}_{0} &\longrightarrow \ \Omega ^{1}(\partial D,\mathfrak{g})\oplus 
\mathbb{C}
\\
\lbrack A] &\longrightarrow\ (A|_{\partial D},1). \label{C}
\end{split}
\end{equation}
This follows from taking $F=0$ in \eqref{equivariance}.

Now, we identify the symplectic leaves in $\mathcal{M}_{0}/H$. Set $\partial D=S^{1}$. The central
extension $\widehat{L\mathfrak{g}}$ of $L\mathfrak{g}$ is defined by the
cocycle \eqref{CS cocycle} above. The adjoint action of $\widehat{L\mathfrak{g}}$ on itself
is given by%
\begin{equation}
ad_{\eta }(\overline{\eta },t)=([\eta ,\overline{\eta }],c(\eta ,\overline{%
\eta }))
\end{equation}%
and this is the infinitesimal version of the adjoint action of the loop
group $LG.$ For $g:S^{1}\rightarrow G,$ consider $g^{-1}\partial _{\sigma }g$
as an element of $\widehat{L\mathfrak{g}},$ i.e. as a mapping $%
S^{1}\rightarrow \mathfrak{g.}$ A bi-linear form on $L\mathfrak{g}$ is
defined by%
\begin{equation}
\left\langle \eta ,\overline{\eta }\right\rangle \longrightarrow \frac{\overline{k}}{2\pi }%
\int\nolimits_{S^{1}}d\sigma \left\langle \eta \overline{\eta }\right\rangle
.
\end{equation}%
Now, the adjoint action of $LG$ is%
\begin{equation}
Ad_{g}(\eta ,t)=(Ad_{g}\eta ,t-\left\langle g^{-1}dg ,
\eta \right\rangle ).
\end{equation}%
Let us describe the co-adjoint action on the subspace $\Omega ^{1}(S^{1},%
\mathfrak{g})\oplus 
\mathbb{C}
$ of  $\widehat{L\mathfrak{g}}^{\ast }.$ Define the pairing between $\Omega
^{1}(S^{1},\mathfrak{g})\oplus 
\mathbb{C}
$ and $\widehat{\mathfrak{h}}$ by (cf. eq. \eqref{pairing})
\begin{equation}
\left\langle (A,z),(\eta ,t)\right\rangle \longrightarrow -\frac{\overline{k}}{%
2\pi }\int\nolimits_{S^{1}}\left\langle \eta A\right\rangle + zt .
\end{equation}%
Then,%
\begin{equation}
\begin{split}
\left\langle Ad_{g}^{\ast }(A,z),(\eta ,t)\right\rangle  &=\left\langle
(A,z),Ad_{g^{-1}}(\eta ,t)\right\rangle  \\
&=-\frac{\overline{k}}{2\pi }\int\nolimits_{S^{1}}d\sigma \left\langle \eta
(Ad_{g}A_{\sigma }-z\partial _{\sigma }gg^{-1})\right\rangle +zt
\end{split} 
\end{equation}%
and
\begin{equation}
Ad_{g}^{\ast }(A,z)=(Ad_{g}A-zdgg^{-1},z). \label{coad action}
\end{equation}
The case of interest is $z=1$, corresponding to the gauge transformations \eqref{gauge} restricted to $S^{1}$. Notice that they are consistent with
the equivariance of the moment map we found above \eqref{C}. Then, the symplectic leaves of \eqref{CS KM} are in one-to-one correspondence with the co-adjoint orbits right above. This is the content of \eqref{isomorphism}.

Consider now the transport matrix for $A\in \Omega ^{1}(S^{1},\mathfrak{g})$
along the disc boundary $S^{1}$    
\begin{equation}
T(A|\sigma _{2},\sigma _{1})=P\exp \left[ -\int\nolimits_{\sigma
_{1}}^{\sigma _{2}}d\sigma A_{\sigma }(\sigma )\right] .
\end{equation}%
Under the co-adjoint action \eqref{coad action} with $z=1$, we have that 
\begin{equation}
T(Ad_{g}^{\ast }A|\sigma _{2},\sigma _{1})=g(\sigma _{2})T(A|\sigma
_{2},\sigma _{1})g(\sigma _{1})^{-1}.
\end{equation}%
Then, the action on the monodromy matrix $m(A)=T(A|2\pi ,0)$ is given by 
\begin{equation}
m(Ad_{g}^{\ast }A)=g(0)m(A)g(0)^{-1}
\end{equation}%
and this shows that the co-adjoint orbits in $\widehat{%
L\mathfrak{g}}^{\ast }$ of the loop group $LG$ are in one-to-one correspondence with the orbits of $%
G$ acting on itself by conjugation. Thus, the symplectic form \eqref{symplectic CS} induces a Poisson structure on $\mathcal{M}$, the symplectic leaves are then obtained by fixing the conjugacy classes of the monodromy matrix $m(A)$ along the boundary $\partial D=S^{1}$. This is precisely the gauge symmetry enhancement we observed before in \eqref{full gauge act}.

\subsection{Hamiltonian structure} \label{3.3}

The Hamiltonian analysis complements the symplectic approach and helps to elucidate further properties of the phase space, like the r\^ole played by the other fields entering the theory or the process of gauge fixing. In this subsection, the relation between the CS and the lambda model action functionals is also considered. We will show how to retrieve from the CS theory, an action functional that is quite close to the original lambda model action functional but ambiguous due to the extension process. Because of such an ambiguity is absent in the case of the lambda deformed PCM, we will consider that situation as a consistency test of the construction and show how to extract the deformed metric and B-field of the deformed PCM from the CS action functional. 

The phase space associated to the Lagrangian \eqref{double CS Lag} is described by the
following data: three conjugate pairs of fields $(A_{i},P_{i}),$ $i=1,2$
and $(A_{\tau },P_{\tau })$ obeying the fundamental Poisson bracket relations\footnote{%
The 2+1 notation used in this section is: $\epsilon_{12}=1$ and $\delta _{xy}$=$\delta^{(2)}(x-y)$.}
\begin{equation}
\{A_{i}(x)_{\mathbf{1}},P_{j}(y)_{\mathbf{2}}\}=\delta _{ij}C_{\mathbf{12}%
}\delta _{xy},\text{ \ \ }\{A_{\tau }(x)_{\mathbf{1}},P_{\tau }(y)_{\mathbf{2%
}}\}=C_{\mathbf{12}}\delta _{xy}
\end{equation}%
and a time evolution determined by the following canonical Hamiltonian (the action is
first order in the time derivative)%
\begin{equation}
h_{C}=-\frac{\overline{k}}{2\pi }\dint\nolimits_{D}\left\langle A_{\tau
}F\right\rangle +\frac{\overline{k}}{4\pi }\dint\nolimits_{\partial
D}\left\langle A_{\tau }A\right\rangle 
\end{equation}%
through the relation%
\begin{equation}
\partial _{\tau }f=\{f,h_{C}\},
\end{equation}%
where $f$ is an arbitrary functional of the phase space variables. 

The definition of the functional derivatives $\delta f/\delta A$ to be used
in the Poisson brackets is subtle because of the presence of
boundaries \cite{Regge-Teitelboim,Banhados}. To find them, we start by computing the variation $\delta f(A)
$ and subsequently investigate under what conditions the result can be
written as an integral over the disc $D$. For example, for the canonical
Hamiltonian, we find that%
\begin{equation}
\delta h_{C}=-\frac{\overline{k}}{2\pi }\dint\nolimits_{D}\left\langle
\delta A_{\tau }F+\delta ADA_{\tau }\right\rangle -\frac{\overline{k}}{4\pi }%
\dint\nolimits_{\partial D}d\sigma \left\langle \delta A_{\sigma }A_{\tau
}-\delta A_{\tau }A_{\sigma }\right\rangle .\label{var hc}
\end{equation}%
Then, in order to cancel the boundary contribution we impose the boundary
equations of motion \eqref{Boundary EOM} and obtain the variation
\begin{equation}
\delta h_{C}=-\frac{\overline{k}}{2\pi }\dint\nolimits_{D}\left\langle
\delta A_{\tau }F+\delta ADA_{\tau }\right\rangle . \label{var hc 2}
\end{equation}

Now, we are ready to consider the Dirac procedure. There are three primary
constraints given by
\begin{equation}
\phi _{i}=P_{i}-\frac{\overline{k}}{4\pi }\epsilon _{ij}A_{j}\approx 0,\text{
\ \ }P_{\tau }\approx 0.
\end{equation}%
Using these constraints we construct the total Hamiltonian%
\begin{equation}
h_{T}=h_{C}+\dint\nolimits_{D}d^{2}x\left\langle u_{i}\phi _{i}+u_{\tau
}P_{\tau }\right\rangle ,
\end{equation}%
where $u_{i}$ and $u_{\tau }$ are arbitrary Lagrange multipliers.

Stability of the primary constraint $P_{\tau }\approx 0$ under the time flow
of $h_{T}$ leads to a secondary constraint%
\begin{equation}
F\approx 0, \label{secondary F}
\end{equation}%
which is nothing but the first bulk equation of motion in \eqref{bulk EOM}. This
follows directly from \eqref{var hc 2}. Otherwise, the secondary
constraint would be modified by a boundary contribution altering the
Euler-Lagrange equations of motion as well. Stability of the constraints $\phi
_{i}\approx 0$ does not produce new constraints but instead fix the Lagrange
multipliers to the values%
\begin{equation}
u_{i}\approx D_{i}A_{\tau }. \label{u1}
\end{equation}%
In order to study the time stability of the secondary constraint \eqref{secondary F}, we shall
consider the more general functional $H(\eta )$ defined in \eqref{gauge Hamiltonian} and this is
because it has a well-defined functional variation. Namely,%
\begin{equation}
\delta H(\eta )=\frac{\overline{k}}{2\pi }\dint\nolimits_{D}\left\langle
\delta AD\eta \right\rangle .
\end{equation}%
Having a well-defined functional, we construct the extended Hamiltonian%
\begin{equation}
h_{E}=h_{T}+H(\overline{\eta }), \label{extended H}
\end{equation}%
where $\overline{\eta }$ now plays the r\^ole of an arbitrary Lagrange
multiplier. Running again, we verify that the stability condition for the constraint $%
P_{\tau }\approx 0$ remains the same but the stability conditions for the
constraints $\phi _{i}\approx 0$ leads to the modifications of \eqref{u1}
\begin{equation}
u_{i}\approx D_{i}(A_{\tau }-\overline{\eta }).  \label{mult}
\end{equation}%
Now, the stability of the secondary constraint under the time flow of $h_{E}$
is given by
\begin{equation}
\{H(\eta ),h_{E}\}\approx -\frac{\overline{k}}{2\pi }\dint\nolimits_{D}\left%
\langle [\eta ,A_{\tau }-\overline{\eta }]F\right\rangle \approx 0,
\end{equation}%
where we have used \eqref{mult} and imposed the condition $\eta |_{\partial D}=0$ on the test functions in
order to reach the final form. For these kind of test functions we denote
the smeared second class constraint by $H_{0}(\eta )$ and from this we conclude that
\begin{equation}
\{H_{0}(\eta ),h_{E}\}\approx 0.
\end{equation}%
Then, the secondary constraint is stable (under any of the flows of $h_{E}$, $h_{T}$ or $h_{C}$) only for those gauge parameters $\eta \in \mathbf{g}_{0}$ in the Lie algebra of the normal subgroup $\mathcal{G}_{0}$ defined in \eqref{normal subgroup} corresponding to the
gauge symmetry group of the CS action functional, see \eqref{gauge conditions}. There are no tertiary constraints produced at this stage and the algorithm stops.

We now find the first class and second class constraints. There is a primary first
class constraint 
\begin{equation}
P_{\tau }\approx 0
\end{equation}%
and two primary second class constraints 
\begin{equation}
\phi _{i}\approx 0.
\end{equation}%
There is another first class constraint formed from a mixture of primary and secondary constraints that is given by
\begin{equation}
\overline{H}_{0}(\eta )=H_{0}(\eta )+\dint\nolimits_{D}d^{2}x\left\langle \eta D_{i}\phi
_{i}\right\rangle ,
\end{equation}
where $\eta \in \mathbf{g}_{0}$. The last condition on the gauge parameters guarantee that the Poisson bracket of two such constraints vanish weakly. 

It is convenient to impose the second class secondary constraints strongly through a
Dirac bracket. After doing this, the phase space of the CS theory is now described by the brackets
\begin{equation}
\{A_{i}(x)_{\mathbf{1}},A_{j}(y)_{\mathbf{2}}\}^{\ast}=\frac{2\pi }{\overline{k}}\epsilon _{ij}C_{\mathbf{12}%
}\delta _{xy},\text{ \ \ }\{A_{\tau }(x)_{\mathbf{1}},P_{\tau }(y)_{\mathbf{2%
}}\}^{\ast}=C_{\mathbf{12}}\delta _{xy} \label{2d Gauge field PB}
\end{equation}%
plus the remaining two first class constraints
\begin{equation}
P_{\tau}\approx 0,\text{ \ \ } H_{0}(\eta)\approx 0.
\end{equation}
The time evolution is given by \eqref{extended H} with $\phi_{i}=0$ and under the time flow of such $h_{E}$, one can show that the second equation of motion in \eqref{bulk EOM} extends to
\begin{equation}
\partial _{\tau }A=\{A,h_{E}\}=D(A_{\tau }-\overline{\eta }).  \label{ham A eom}
\end{equation}
For arbitrary functionals of $A_{i}$ defined on $D$, the first bracket in \eqref{2d Gauge field PB} generalize to\footnote{%
For arbitrary functions of $A_{\tau },$ the Poisson bracket is obvious and
will not be written.}%
\begin{equation}
\{f,g\}^{\ast}(A) = \int\nolimits_{D}d^{2}xd^{2}y\frac{\delta f(A)}{\delta A_{i}^{A}(x)}%
\{A_{i}^{A}(x),A_{j}^{B}(y)\}^{\ast}\frac{\delta g(A)}{\delta A_{j}^{B}(y)},  \label{CS PB 0}
\end{equation}%
or equivalently,
\begin{eqnarray}
\{f,g\}^{\ast}(A) &=&\frac{2\pi }{\overline{k}}\epsilon _{ij}\int\nolimits_{D}d^{2}x\frac{%
\delta f(A)}{\delta A_{i}^{A}(x)}\eta ^{AB}\frac{\delta g(A)}{\delta
A_{j}^{B}(x)}. \label{CS PB}
\end{eqnarray}%
In what follows we will omit the symbol $\ast$ in order to avoid clutter and refer to these Dirac brackets simply as the CS Poisson brackets. This is to match the literature where the brackets \eqref{2d Gauge field PB} are usually taken as the defining CS Poisson brackets.

The Poisson algebra of $H(\eta)$ computed directly from \eqref{CS PB} matches \eqref{gauge Ham PB} perfectly and we recover the first class Poisson sub-algebra by restricting to those gauge parameters $\eta \in \mathbf{g}_{0}$, as expected. The action functional itself is not invariant under the action of the gauge group $\mathcal{G}/\mathcal{G}_{0}$.
However, this group is a symmetry of the phase space as the
latter is identified with the space of solutions to the equations of motion which often have a more general symmetry structure. The action of $\mathcal{G}/\mathcal{G}_{0}$ is a perfectly well-defined symmetry of the boundary theory (recall $F=0$ is an equation of motion and a constraint) as shown above by using the symplectic approach. 

We now consider gauge fixing and exploit the fact that Dirac brackets can be imposed at stages, i.e. we
gauge fix each first class constraint at the time and then impose the
associated Dirac bracket independently of the other constraints. In order to
take $P_{\tau }=0$ strongly, we choose the following gauge fixing condition%
\begin{equation}
A_{\tau }-\mathcal{L}_{\tau }\approx 0,  \label{P fix}
\end{equation}%
where $\mathcal{L}_{\tau }=\mathcal{L}_{\tau }(A)$ is a function of the
two components of the gauge field $A$, whose explicit form is not
relevant in what follows. The only property $\mathcal{L}_{\tau
}(A)$ must have is that at $\partial D$ the boundary equations of
motion \eqref{Boundary EOM} are satisfied (recall that our previous results depends of this hypothesis). The choice \eqref{P fix} is a good gauge fixing condition and its stability under
the flow of $h_{E}$ leads to a condition on the so far unspecified Lagrange
multiplier
\begin{equation}
u_{\tau }\approx \{\mathcal{L}_{\tau },h_{E}\},
\end{equation}%
whose explicit form is not required at this point because it couples with $P_{\tau }$ in $h_{E}$ and drops out at the end anyway. We discard the second PB in \eqref{2d Gauge field PB}, while continue using the first one because it is not modified under this gauge fixing. 

The time evolution is now determined by the Hamiltonian
\begin{equation}
h_{E}=h_{C}+H(\overline{\eta}) \label{partial h_E}
\end{equation}
and its time flow induce an extended Hamiltonian equation of motion. Indeed, if we write in general
\begin{equation}
A_{\tau }=\mathcal{O}_{2}^{T}A_{1}-\mathcal{O}_{1}^{T}A_{2}, \label{O's}
\end{equation}%
where $\mathcal{O}_{i}$ are projectors along the $%
\mathbb{Z}
_{4}$ grading of $\mathfrak{f}$, the equations of motion \eqref{ham A eom} take the form%
\begin{equation}
\partial _{\tau }A_{i}=\{A_{i},h_{E}\}=D_{i}(A_{\tau }-\overline{\eta })+%
\mathcal{O}_{i}F_{12}. \label{ext ham A eom}
\end{equation}%
Clearly, they are extensions of the eom by terms proportional to the first class
constraint $F\approx 0$.

The new ingredient of the Hamiltonian approach is the presence of the conjugate pair $(A_{\tau},P_{\tau})$, which is not taken into account in the symplectic approach. The symplectic form associated to the first CS Poisson bracket\footnote{When inverted on the symplectic leaves.} in \eqref{2d Gauge field PB} is the two form \eqref{symplectic CS}. For instance, we can write \eqref{CS PB} in several standard ways
\begin{equation}
\{f,g\}(A)=\omega (X_{\eta },X_{\xi })=X_{\eta }(g)=\delta g(X_{\eta
})=-X_{\xi }(f)=-\delta f(X_{\xi }),
\end{equation}%
where $X_{\eta }$ and $X_{\xi }$ are the Hamiltonian vector fields
associated to the functionals $f$ and $g$, respectively. The relation between the symplectic and Poisson structures being%
\begin{equation}
\omega (X_{\eta },X_{\xi })=\frac{\overline{k}}{2\pi }\dint\nolimits_{D}%
\left\langle \eta \wedge \xi \right\rangle ,
\end{equation}%
where 
\begin{equation}
\eta =\eta _{i}dx^{i},\text{ \ \ }X_{\eta }=\eta _{i}\frac{\delta }{\delta
A_{i}},\text{ \ \ }\eta _{i}=\frac{2\pi }{\overline{k}}\frac{\delta f}{%
\delta A_{i}}
\end{equation}%
and similar expressions for the pair $(\xi ,g)$. The connection between both approaches is now clear.

The only remaining first class constraint is $H_{0}(\eta)\approx 0$. However, instead of gauge fixing it we invoke the results concerning the symplectic reduction presented in \eqref{3.2}. In this case the restriction of the CS Poisson bracket to the constraint surface holds through its symplectic form and gives rise to the Kac-Moody boundary algebra
\begin{equation}
\{A_{i}(x)_{\mathbf{1}},A_{j}(y)_{\mathbf{2}}\}_{CS}\longrightarrow \{A_{\sigma }(\sigma )_{\mathbf{1}},A_{\sigma }(\sigma ^{\prime
})_{\mathbf{2}}\}_{KM}. 
\end{equation}
 
The contact with the lambda models is now obvious, at $\partial D$, we make the following identifications
\begin{equation}
A_{\sigma (\pm )}=\overline{\mathscr{L}}_{\sigma }(z_{\pm }),\text{ \ \ }A_{\tau (\pm
)}=\overline{\mathscr{L}}_{\tau }(z_{\pm }). \label{pole identification}
\end{equation}%
The last condition ensures that the CS boundary equations of motion are satisfied.
Notice that%
\begin{equation}
A_{\tau (\pm )}|_{\partial D}=\mathcal{L}_{\tau (\pm )}(A_{i})|_{\partial D}=%
\overline{\mathscr{L}}_{\tau }(z_{\pm }).
\end{equation}
This condition might be used to fix the explicit forms of the projectors $\mathcal{O}_{i}$ in \eqref{O's} if wanted.

Once we have symplectic-reduced the CS theory the last thing to understand is the
time evolution equation in terms of the boundary Poisson structure. Indeed, as expected, we find that
\begin{equation}
\partial _{\tau }A_{\sigma }=\{A_{\sigma },h_{E}\}_{KM},
\end{equation}%
where%
\begin{equation}
h_{E}=\frac{\overline{k}}{4\pi }\int\nolimits_{\partial D}d\sigma
\left\langle A_{\tau }A_{\sigma }\right\rangle -\frac{\overline{k}}{2\pi }%
\int\nolimits_{\partial D}d\sigma \left\langle \overline{\eta }A_{\sigma
}\right\rangle .
\end{equation}
Notice that $h_{E}|_{\overline{\eta }=0}$ is precisely the extended
lambda model Hamiltonian $\overline{h}$ we constructed before in \eqref{extended H,P} for the lambda model. 
With the help of \eqref{p on J} and \eqref{j and l} we verify that%
\begin{equation}
\partial _{\tau }A_{\sigma }=D_{\sigma }(A_{\tau }-\overline{\eta }),
\end{equation}%
in consistency with \eqref{ham A eom} and \eqref{ext ham A eom} when restricted to $\partial D$. For $\overline{\eta }=0$, we do recover \eqref{strong flatness} at the points $z=z_{\pm}$. The generator of translations $\overline{p}$ along the boundary $\partial D$, follows directly from the expression \eqref{extended H,P} as well.

The boundary gauge symmetry takes the expected form 
\begin{equation}
\delta _{\eta }A_{\sigma }=\{A_{\sigma },H(\eta )\}_{KM}=-D_{\sigma }\eta ,
\end{equation}%
where we have used the generator \eqref{bdry gauge Ham}. Its exponentiation $g=\exp \eta $ is precisely \eqref{coad action} with $z=1$ at $\partial D$
\begin{equation}
Ad_{g}^{\ast }A_{\sigma }=gA_{\sigma }g^{-1}-\partial _{\sigma }gg^{-1}.
\end{equation}
Then,
\begin{equation}
m^{\prime }(z_{\pm })=m(Ad_{g_{\pm }}^{\ast }A).
\end{equation}
The conclusion is that the phase space of the symplectic reduced double Chern-Simons action is equivalent to the phase space of the extension of the lambda model. As shown before, the gauge fixing of the conjugacy classes associated to the first class constraints of the lambda model that leads to the true physical degrees of freedom is accomplished through the used of the dressing gauge \eqref{dressing gauge}.

Now, we consider the relation between the action functionals of the Chern-Simons theory and the lambda
model. In order to see it we compute the effective
action induced by the use of the eom $F_{(\pm )}=0$ in the CS theory. In this case, the phase space Poisson
structure is that of the Kac-Moody algebra and the pair $(A_{\tau
},A_{\sigma })$ constitute a flat gauge connection. Hence, we write   
\begin{equation}
A_{\tau (\pm )}=-\partial _{\tau }\overline{\Psi }(z_{\pm })\overline{\Psi }%
(z_{\pm })^{-1},\text{ \ \ }A_{\sigma (\pm )}=-\partial _{\sigma }\overline{%
\Psi }(z_{\pm })\overline{\Psi }(z_{\pm })^{-1},\text{ \ \ on \ \ }\partial
D,  \label{gauge-lax}
\end{equation}%
because of the identifications \eqref{pole identification}. On the
other hand, we have that 
\begin{equation}
A_{i(\pm )}=-\partial _{i}\overline{\Psi }^{\prime }(z_{\pm })\overline{\Psi 
}^{\prime }(z_{\pm })^{-1}\text{ \ \ on \ \ }D,
\end{equation}%
where the prime is to emphasize that  $\overline{\Psi }^{\prime }$ is an
extension of $\overline{\Psi }$ from $\partial D$ to $D$. By writing the
action \eqref{action on DxR} in the form $S=S_{D}+S_{\partial D}$%
, where $S_{D}$ is given by the first term in the rhs while $S_{\partial D}$
by the second one, we obtain%
\begin{equation}
S_{D} =\frac{\overline{k}}{4\pi }\int\nolimits_{\partial D\times 
\mathbb{R}
}d^{2}\sigma \big\langle \partial _{\tau }\overline{\Psi }\overline{%
\Psi }^{-1}\partial _{\sigma }\overline{\Psi }\overline{\Psi }%
^{-1}\big\rangle +\frac{\overline{k}}{4\pi }\int\nolimits_{D\times 
\mathbb{R}
}\chi (\overline{\Psi }') \label{effective-disc}
\end{equation}
and
\begin{equation}
S_{\partial D} =-\int\nolimits_{%
\mathbb{R}
}d\tau \overline{h},
\end{equation}
where we have used \eqref{extended H,P}. By the first relation in \eqref{gauge-lax} and the fact that $A_{\tau (\pm )}$ is fixed at $%
\partial D,$ it turns out that the kinetic term contribution to the action in \eqref{effective-disc} is precisely $-S_{\partial D}$ and the resulting
effective action does not seem to have any boundary contributions\footnote{%
At this point is worth it to recall the result \eqref{action on-shell} and the comment below.}. However, in terms of the field
variable 
\begin{equation}
\overline{\mathcal{F}}=\overline{\Psi }(z_{+})\overline{\Psi }(z_{-})^{-1},
\end{equation}%
the properties of the WZ term imply that%
\begin{equation}
S_{eff}=\frac{k}{2\pi }\int\nolimits_{\partial D\times 
\mathbb{R}
}d^{2}\sigma \big\langle \overline{\mathcal{F}}^{-1}\partial _{-}\overline{\mathcal{F}}\overline{%
\mathscr{L}}_{+}(z_{-})-\overline{\mathcal{F}}^{-1}\partial _{+}\overline{\mathcal{F}}\overline{%
\mathscr{L}}_{-}(z_{-})\big\rangle +\frac{k}{4\pi }\int\nolimits_{D\times 
\mathbb{R}
}\chi (\mathcal{\overline{F}'}), \label{effective action general}
\end{equation}%
where
\begin{equation}
\overline{\mathscr{L}}_{\pm }(z_{-})=-\partial _{\pm }%
\overline{\Psi }(z_{-})\overline{\Psi }(z_{-})^{-1}
\end{equation}
is identified with the extended
Lax connection \eqref{extended Light-cone Lax}. The boundary
effective action obtained from the CS theory is that of the extended lambda model in the partial
gauge considered in section \eqref{2.2}.

Let us now seek for a more manifest relation between the actions \eqref{deformed-GS} and \eqref{effective action general}. We will start
by writing the original lambda model action \eqref{deformed-GS} in terms of the constraints
\eqref{secondary const}. After some algebra, we find 
\begin{equation}
\begin{aligned}
S_{\lambda } =&-\frac{k}{2\pi }\int\nolimits_{\Sigma }d^{2}\sigma
\left\langle \mathcal{F}^{-1}\partial _{-}\mathcal{F}(\Omega ^{T}A_{+}+(2\pi
/k)C_{+}\mathcal{)-F}^{-1}\partial _{+}\mathcal{F}A_{-}\right\rangle  \\
&-\frac{k}{4\pi }\int\nolimits_{B}\chi (\mathcal{F})+\int\nolimits_{\Sigma
}d^{2}\sigma \left\langle A_{+}C_{-}+A_{-}C_{+}\right\rangle .
\end{aligned}
\end{equation}%
Now, in the partial gauge \eqref{cons c minus}, \eqref{partial} and with the help of \eqref{phi const} and \eqref{extended Light-cone Lax}, we get%
\begin{equation}
S_{\lambda }=-S_{eff}+\frac{k}{2\pi }\int\nolimits_{\Sigma }d^{2}\sigma
\big\langle \alpha z_{+}^{4}(\varphi ^{(1)}+\varphi ^{(3)})(\mathcal{F}%
^{-1}\partial _{-}\mathcal{F-F}^{-1}\partial _{+}\mathcal{F})+(2\pi
/k)A_{-}^{(3)}\varphi ^{(1)}\big\rangle .
\end{equation}%
Both actions are not the same probably because of in the whole Hamiltonian analysis discussed so far, we have
not specified how the original lambda model Lagrangian field $\mathcal{F}$ extends
outside the constraint surface\footnote{Indeed, one of the actions is written in terms of $\overline{\mathcal{F}}$, while the other in terms of $\mathcal{F}$.} in such a way that its Euler-Lagrange
equations of motion derived from \eqref{effective action general} produce two flat connections at the poles $z_{\pm}$. What is important here is
that both actions coincide on the constraint surface. Notice that the discrepancy is linked purely to the fermionic constraints, perhaps related to the choice made in \eqref{F}.
However, after gauge fixing (dressing gauge, etc) both actions functionals describe the
same physical system, as expected. In any case, the strongly flatness of the extended Lax connection can be traced back to originate from the CS\ theory. The sign of both actions can be matched by reversing the super-trace sign in the CS action \eqref{action on DxR} or by taking $S_{CS}\rightarrow -S_{CS}$. 

As a particular simple an unambiguous example, let us consider the lambda deformed PCM \cite{Sfetsos}. In this case all the constraints are second class \cite{lambda-bos}, the original and the extended
Lax connection coincide so the Lax connection is already strongly flat \cite{in progress} and this means that the relations \eqref{solutions-wave} can be used inside the
action functional. We get,
\begin{equation}
S_{eff}=\frac{k}{2\pi }\int\nolimits_{\partial D\times 
\mathbb{R}
}d^{2}\sigma \left\langle %
\mathcal{F}^{-1}\partial _{-}\mathcal{F}\Omega ^{T}A_{+}-\mathcal{F}^{-1}\partial _{+}\mathcal{F}A_{-}\right\rangle +\frac{%
k}{4\pi }\int\nolimits_{D\times 
\mathbb{R}
}\chi (\mathcal{F'}),
\end{equation}%
where the $A_{\pm }$ depends on $\mathcal{F}$ via \eqref{gauge field eom} and $\Omega =\lambda ^{-1}$ is the omega projector characteristic
of the PCM. After some algebraic manipulations, we obtain%
\begin{equation}
S_{eff}=\frac{k}{2\pi }\int\nolimits_{\partial D\times 
\mathbb{R}
}d^{2}\sigma \left\langle \mathcal{F}^{-1}\partial _{+}\mathcal{F}(G+B)%
\mathcal{F}^{-1}\partial _{-}\mathcal{F}\right\rangle, \label{effective}
\end{equation}%
where%
\begin{equation}
\begin{aligned}
G &=\frac{1}{(\Omega -D)}(\Omega \Omega ^{T}-1)\frac{1}{(\Omega ^{T}-D^{T})}%
, \\
B &=B_{0}+\frac{1}{(\Omega -D)}(D\Omega ^{T}-\Omega D^{T})\frac{1}{(\Omega
^{T}-D^{T})},
\end{aligned}
\end{equation}%
are associated to the deformed background metric and antisymmetric field with $B_{0}$ representing the contribution from the WZ term.
The effective action \eqref{effective} is (up to a global sign) the action functional of the lambda deformed
PCM \cite{Sfetsos} obtained from the action \eqref{deformed-GS} after integration of the gauge fields $A_{\pm}$. 

\subsection{The spectral parameter} \label{3.4}

We now introduce the dependence of the spectral parameter $z$ in the CS theory that is behind the integrable structure of the lambda models. The most important result in this subsection is the relation between a $z$-dependent extension of the Goldman bracket and the would-be Poisson algebra of the lambda deformed monodromy matrices under the symplectic reduction process of the CS theory. 

Make use of the twisted loop superalgebra structure \eqref{loop superalgebra} and consider the problem of finding a $z$-dependent
2-dimensional gauge field $A(z)$ on the disc $D$ satisfying the following two conditions
\begin{equation}
A(z_{\pm})=A_{(\pm )}\text{ \ \ and \ \ }\Phi (A(z))=A(iz). \label{conditions}
\end{equation}%
The answer we will consider here (recall that $A(z)=A_{i}(z)dx^{i}$, $i=1,2$) is given by
\begin{equation}
A(z)=-\frac{k}{2\pi}f_{-}(z)\overline{\Omega }(z/z_{+})A_{(+)}+\frac{k}{2\pi}f_{+}(z)\overline{%
\Omega }(z/z_{-})A_{(-)},
\label{interpolating current}
\end{equation}%
with $f_{\pm}(z)$ and $\overline{\Omega }(z)$defined by \eqref{f's} and \eqref{Omega Hybrid}, respectively. 

Using this $z$-dependent gauge field we gather both (for $\overline{k}=\pm k$) Poisson brackets corresponding to the first expression in (\ref{2d Gauge
field PB}) into a single interpolating one
\begin{equation}
\{A_{i}(x,z)_{\mathbf{1}},A_{j}(y,w)_{\mathbf{2}}\}=-2\mathfrak{s}%
_{\mathbf{12}}(z,w)\epsilon _{ij}\delta _{xy},  \label{pre Maillet}
\end{equation}%
which is the precursor of the Maillet bracket as we shall see. We have that
\begin{equation}
\mathfrak{s}_{\mathbf{12}}(z,w)=-\frac{k}{4\pi }[f_{-}(z)f_{-}(w)\overline{%
\Omega }(z/z_{+})_{\mathbf{1}}\overline{\Omega }(w/z_{+})_{\mathbf{2}%
}-f_{+}(z)f_{+}(w)\overline{\Omega }(z/z_{-})_{\mathbf{1}}\overline{\Omega }%
(w/z_{-})_{\mathbf{2}}]C_{\mathbf{12}} \label{proj s}
\end{equation}
turns out to be the same \cite{hybrid} as the $\mathfrak{s}%
_{\mathbf{12}}(z,w)$ appearing in the Maillet bracket \eqref{Maillet-lambda} of the lambda model and equivalently written in the second line of \eqref{def r,s}. Notice that the theory \eqref{CS copies} actually consist of two Chern-Simons theories with opposite levels attached to the poles $z_{\pm}$ of \eqref{twisting function} in the complex plane. At this stage, we assume we are already impose the partial gauge fixing conditions
\begin{equation}
P_{\tau }\approx 0,\text{ \ \ }A_{\tau }\approx \mathcal{L}_{\tau }
\end{equation}
strongly, where the second Poisson bracket in \eqref{2d Gauge
field PB} is absent.

We now compute the $z$-dependent
boundary algebra that results after performing the symplectic reduction \eqref{isomorphism}. We take $i=\sigma$ in \eqref{interpolating current} and use the KM algebra \eqref{CS KM}. 
The boundary $z$-dependent Poisson algebra is nothing but the Maillet algebra
\begin{equation}
\begin{aligned}
\{A_{\sigma}(\sigma ,z)_{\mathbf{1}},A_{\sigma}(\sigma
^{\prime },w)_{\mathbf{2}}\}= \: &[R_{\mathbf{12}}(z,w),A_{\sigma }(\sigma ,z)_{\mathbf{1}}]\delta _{\sigma
\sigma ^{\prime }}-[R_{\mathbf{12}}^{\ast }(z,w),A_{\sigma }(\sigma ^{\prime
},w)_{\mathbf{2}}]\delta _{\sigma \sigma ^{\prime }} \\
&-(R_{\mathbf{12}}(z,w)+R_{\mathbf{12}}^{\ast }(z,w))\delta
_{\sigma \sigma ^{\prime }}^{\prime }, \label{Maillet}
\end{aligned}
\end{equation}%
where%
\begin{equation}
\begin{aligned}
R_{\mathbf{12}}(z,w)&=-\frac{2}{z^{4}-w^{4}}\tsum%
\nolimits_{j=0}^{3}z^{j}w^{4-j}C_{\mathbf{12}}^{(j,4-j)}\varphi _{\lambda }^{-1}(w),\\
R_{\mathbf{12}}^{\ast }(z,w)&=R_{\mathbf{21}}(w,z)=\frac{2}{z^{4}-w^{4}}\tsum%
\nolimits_{j=0}^{3}w^{j}z^{4-j}C_{\mathbf{12}}^{(4-j,j)}\varphi _{\lambda }^{-1}(z). \label{R12}
\end{aligned}
\end{equation}
Alternatively, by setting 
\begin{equation}
\mathfrak{r}_{\mathbf{12}}(z,w)=\frac{1}{2}(R_{\mathbf{12}}(z,w)-R_{\mathbf{12}}^{\ast }(z,w))
\end{equation}%
and
\begin{equation}
\mathfrak{s}%
_{\mathbf{12}}(z,w)=\frac{1}{2}(R_{\mathbf{12}}(z,w)+R_{\mathbf{12}}^{\ast }(z,w)),
\end{equation}
we recover \eqref{Maillet-lambda} after identifying
\begin{equation}
A_{\sigma}(\sigma;z)=\overline{\mathscr{L}}_{\sigma}(\sigma;z). \label{key rel 1}
\end{equation}
The picture is completed by introducing and interpolating field satisfying the second relations in \eqref{pole identification}. The obvious answer is
\begin{equation}
A_{\tau}(\sigma;z)=\overline{\mathscr{L}}_{\tau}(\sigma;z) \label{key rel 2}
\end{equation} 
and a $z$-dependent flat connection $F_{\tau \sigma}(z)=0$ is automatically constructed by virtue of the Kac-Moody algebra structure of the boundary theory. Clearly, this is the strongly flat extended Lax connection constructed in \eqref{2.3}.

In order to consider the Poisson bracket of transport matrices, It is useful
to adapt the bracket \eqref{CS PB 0} to the situation of functionals of the form $%
f(A(x(t)))$, where $x^{i}(t)\subset D$ is any path defined on the disc.
Then, \eqref{CS PB 0} becomes 
\begin{equation}
\{f,g\}(A)=\int dtds\frac{\delta f(A)}{\delta A_{i}(x(t))_{ab}}%
\{A_{i}(x(t))_{ab},A_{j}(y(s))_{cd}\}\frac{\delta g(A)}{\delta
A_{j}(y(s))_{cd}}, \label{CS PB on curve}
\end{equation}%
where we have written the gauge field in an arbitrary matrix representation. 

The transport matrix associated to \eqref{interpolating current} along the path $x^{i}(t^{\prime })$, $%
t^{\prime }\subset \lbrack \overline{t},t]$ is%
\begin{equation}
T(t,\overline{t};z)=P\exp [-\int\nolimits_{\overline{t}}^{t}dt^{\prime }%
\frac{dx^{i}(t^{\prime })}{dt^{\prime }}A_{i}(x(t^{\prime });z)]. \label{transport in CS}
\end{equation}%
Because of the normalization condition $T(t,\overline{t};z)|_{t=\overline{t}}=I$ holds, its variation is given by%
\begin{equation}
\delta T(t,\overline{t};z)=-\int\nolimits_{\overline{t}}^{t}dt^{\prime }%
\frac{dx^{i}(t^{\prime })}{dt^{\prime }}T(t,t^{\prime };z)\delta
A_{i}(x(t^{\prime });z)T(t^{\prime },\overline{t};z).
\end{equation}%
By expressing the variation of $A_{i}(x(t);z)$ in terms of the variations of 
$A_{i(\pm )}(x(t))$ at the poles and using the brackets \eqref{CS PB on curve}, we can compute the Poisson
bracket of two transport matrices associated to the paths $x^{i}(t^{\prime
})$, $t^{\prime }\subset \lbrack \overline{t},t]$  and $y^{i}(s^{\prime })$, 
$s^{\prime }\subset \lbrack \overline{s},s]$. We find
\begin{equation}
\begin{aligned}
\{T(t,\overline{t};z)_{\mathbf{1}},T(s,\overline{s};w)_{\mathbf{2}}\}
=&\int\nolimits_{\overline{t}}^{t}dt^{\prime }\int\nolimits_{\overline{s}%
}^{s}ds^{\prime }\frac{dx^{i}(t^{\prime })}{dt^{\prime }}\frac{%
dy^{j}(s^{\prime })}{ds^{\prime }}  
T(t,t^{\prime };z)_{\mathbf{1}}T(s,s^{\prime };w)_{\mathbf{2}} \\
&\times\{A_{i}(x(t^{\prime });z)_{\mathbf{1}},A_{j}(y(s^{\prime });w)_{\mathbf{2}}\}T
(t^{\prime },\overline{t};z)_{\mathbf{1}}T(s^{\prime },\overline{s};w)_{\mathbf{2}}.
\end{aligned}
\end{equation}%

It is important to mention that we started considering $A_{i(\pm )}$ as the true
phase space variables. The final result above, however, show that
functional variations can be done directly with respect the whole $z$-dependent field $%
A_{i}(z)$ as should be in consistency with \eqref{pre Maillet}. Now, using \eqref{pre Maillet}, we get\footnote{%
Recall that $dx^{i}dy^{j}\epsilon _{ij}=|d\mathbf{x}\times d\mathbf{y}|$
with $\mathbf{x}=(0,x^{1},x^{2})$ and $\mathbf{y}=(0,y^{1},y^{2})$ is the
area 2-form $d^{2}x.$}
\begin{equation}
\begin{aligned}
\{T(t,\overline{t};z)_{\mathbf{1}},T(s,\overline{s};w)_{\mathbf{2}}\}
=&-2\int\nolimits_{\overline{t}}^{t}dt^{\prime }\int\nolimits_{\overline{s}%
}^{s}ds^{\prime }\frac{dx^{i}(t^{\prime })}{dt^{\prime }}\frac{%
dy^{j}(s^{\prime })}{ds^{\prime }}\epsilon _{ij}\delta _{x^{1}(t^{\prime
})y^{1}(s^{\prime })}\delta _{x^{2}(t^{\prime })y^{2}(s^{\prime })}  \\
&\times T(t,t^{\prime };z)_{\mathbf{1}}T(s,s^{\prime };w)_{\mathbf{2}}\mathfrak{s}%
_{\mathbf{12}}(z,w)T(t^{\prime },\overline{t};z)_{\mathbf{1}}T(s^{\prime
},\overline{s};w)_{\mathbf{2}}.
\end{aligned}
\end{equation}%

If the two paths intersect transversely at a single point $p$, then $x^{i}(\hat{s})=y^{i}(%
\hat{s})$ for some $\hat{s}$ and we have%
\begin{equation}
\{T(t,\overline{t};z)_{\mathbf{1}},T(s,\overline{s};w)_{\mathbf{2}}\}=-2%
T(t,\hat{s};z)_{\mathbf{1}}T(s,\hat{s};w)_{\mathbf{2}}\mathfrak{s}%
_{\mathbf{12}}(z,w)T(\hat{s},\overline{t};z)_{\mathbf{1}}T(\hat{s%
},\overline{s};w)_{\mathbf{2}}. \label{single intersec}
\end{equation}%
If the paths intersect more than once then we sum over the discrete set of
contributions. 

From \eqref{transport in CS} we introduce a $z$-dependent Wilson loop on
a closed path $\gamma \subset D$ by identifying the initial and final points $x^{i}(t)=x^{i}(\overline{t}%
)$, 
\begin{equation}
W(\gamma ;z)=P\exp [-\oint\nolimits_{\gamma }dx^{i}A_{i}(x;z)]. \label{Wilson z}
\end{equation}%
Using \eqref{single intersec} for several intersections at the points $p_{n}$, we can compute the Poisson bracket of two Wilson loops   
\begin{equation}
\begin{aligned}
\{W(\gamma _{1};z)_{\mathbf{1}},W(\gamma _{2};w)_{\mathbf{2}}\}
&=&-2\sum\limits_{p_{n}\in \gamma _{1}\cap \gamma _{2}}T(x(t),x(\hat{s}%
_{n});z)_{\mathbf{1}}T(y(s),x(\hat{s}_{n});w)_{\mathbf{2}} \\
&&\times \mathfrak{s}_{\mathbf{12}}(z,w)T(x(\hat{s}_{n}),x(t);z)_{%
\mathbf{1}}T(x(\hat{s}_{n}),y(s);w)_{\mathbf{2}},
\end{aligned} \label{pre monodromy alg}
\end{equation}
where we have emphasized on the identification of the initial and final
points in the rhs. 

The result above suggest introducing a notation in order to write the rhs in a more compact form. 
Let us denote by $W(\gamma^{s}_{*};z)$ the Wilson loop with an insertion $\ast$ at the point $x^{i}(s)$ in the closed path $\gamma$ with base point $x^{i}(t)$, i.e.
\begin{equation}
W(\gamma^{s}_{*};z)=T(x(t),x(s);z)\overset{\downarrow }{\ast }T(x(s),x(t);z).
\end{equation}%
Then, if we decompose $\mathfrak{s}_{\mathbf{12}}(z,w)$ in the form 
\begin{equation}
\mathfrak{s}_{\mathbf{12}}(z,w)=\tsum%
\nolimits_{j=0}^{3}a^{(j)}(z,w)\otimes \overline{a}^{(j)}(z,w),
\end{equation}
we obtain the quadratic algebra
\begin{equation}
\{W(\gamma_{1} ;z)_{\mathbf{1}},W(\gamma_{2} ;w)_{\mathbf{2}}\}= -2\sum\limits_{p_{n}\in \gamma _{1}\cap \gamma _{2}}
W(\gamma^{\hat{s}_{n}}_{1a};z)_{\mathbf{1}}W(\gamma^{\hat{s}_{n}}_{2\overline{a}};w)_{\mathbf{2}}, \label{pre-monodromy}
\end{equation} 
where we have dropped the $\mathbb{Z}_{4}$ index $j$ sum in order to simplify the notation. 

By taking supertraces of \eqref{pre monodromy alg} and using the cyclicity property, we get an important result
\begin{equation}
\{\left\langle W(\gamma _{1};z)\right\rangle ,\left\langle W(\gamma
_{2};w)\right\rangle \}=-2\sum\limits_{p_{n}\in \gamma _{1}\cap \gamma
_{2}}\left\langle W(\gamma _{1}(\hat{s}_{n});z)_{\mathbf{1}}W(\gamma
_{2}(\hat{s}_{n});w)_{\mathbf{2}}\mathfrak{s}_{\mathbf{12}}(z,w)\right\rangle _{%
\mathbf{12}}. \label{z-Goldman}
\end{equation}%
The $W(\gamma _{1}(\hat{s}_{n});z)$ and $W(\gamma _{2}(\hat{s}%
_{n});w)$ above denote two Wilson loops along the closed paths $\gamma _{1}$ and $%
\gamma _{2}$ just like original $W(\gamma _{1};z)$ and $W(\gamma _{2};w)$
but with coinciding base point $x^{i}(\hat{s}_{n})$ and $y^{i}(\hat{s}_{n})$ where the paths intersect. As a particular case, consider \eqref{z-Goldman} evaluated
at the poles $z=z_{\pm }$ of the twisting function. Namely,
\begin{equation}
\{\left\langle W(\gamma _{1})\right\rangle ,\left\langle W(\gamma
_{2})\right\rangle \}=\frac{2\pi }{\overline{k}}\sum\limits_{p_{n}\in \gamma
_{1}\cap \gamma _{2}}\left\langle W(\gamma _{1}(\hat{s}_{n}))_{\mathbf{1}%
}W(\gamma _{2}(\hat{s}_{n}))_{\mathbf{2}}C_{\mathbf{12}}\right\rangle _{\mathbf{12}}.\label{pre-Goldman}
\end{equation}
This last expression is a master equation that is at the core of the Goldman bracket \cite{Goldman}, which is used to study the intersection properties of homotopy classes of loops in genus $g$ Riemann surfaces $\Sigma_{g}$. Depending of the classical Lie algebra, the intersection point is resolved differently by invoking a matrix representation of $C_{\mathbf{12}}$, the final expression involves suitable definitions of the product of loops (i.e. rerouted loops) at the intersection points. For a derivation of \eqref{pre-Goldman} using CS theories as well but in a different context, see \cite{Hassan}. Our construction easily generalizes to the case where the disc is replaced by a genus $g$ Riemann surface $\Sigma_{g}$ with boundaries (recall \cite{Audin}), which is a more natural scenario to be considered due to the novel connection of \eqref{z-Goldman} with the Goldman bracket.


As an example, let us consider the bracket \eqref{z-Goldman} in the simplest possible situation, which corresponds to the lambda model of the PCM and a single intersection point on the torus $\Sigma_{1}$ without boundaries. This is not directly related to our construction because of the absence of boundaries but will help in understanding how \eqref{z-Goldman} works. In this case, the symmetric part of the $R_{\mathbf{12}}(z,w)$ matrix is proportional to the tensor Casimir \cite{quantum-group} and we have
\begin{equation}
\mathfrak{s}_{\mathbf{12}}(z,w)=s(z,w)C_{\mathbf{12}}.  \label{PCM-s}
\end{equation}
Following \cite{Hassan}, we now write the tensor Casimir for several classical Lie groups using a 
$n$-dimensional matrix representation based on the generalized Gell-Mann
matrices. Before continuing, we need to introduce the identity, the
permutation and the defect matrices, denoted respectively by  $I_{\mathbf{12}%
}$, $P_{\mathbf{12}}$ and $\chi _{\mathbf{12}}$. They satisfy the following trace
properties when acting on two arbitrary $n\times n$ matrices $A$ and $B$,
\begin{equation}
\left\langle A_{\mathbf{1}}B_{\mathbf{2}}I_{\mathbf{12}}\right\rangle _{%
\mathbf{12}}=\left\langle A\right\rangle \left\langle B\right\rangle ,\text{
\ }\left\langle A_{\mathbf{1}}B_{\mathbf{2}}P_{\mathbf{12}}\right\rangle
=\left\langle AB\right\rangle ,\text{ \ }\left\langle A_{\mathbf{1}}B_{%
\mathbf{2}}\chi _{\mathbf{12}}\right\rangle =-\big\langle
AB^{-1}\big\rangle .
\end{equation}%
Using these results we have:

\begin{itemize}
\item For $GL(n,%
\mathbb{R}
)$ and $U(n):$%
\begin{equation}
C_{\mathbf{12}}=2P_{\mathbf{12}}
\end{equation}%
and%
\begin{equation}
\{\left\langle W(\gamma _{1};z)\right\rangle ,\left\langle W(\gamma
_{2};w)\right\rangle \}=-4s(z,w)\left\langle W(\gamma _{1}(\hat{s}%
);z)W(\gamma _{2}(\hat{s});w)\right\rangle .  \label{A}
\end{equation}%
At the poles\footnote{For the PCM the dependence of the poles on the parameter $\lambda$ is different \cite{quantum-group}.} $z_{\pm }$,
\begin{equation}
\{\left\langle W(\gamma _{1})\right\rangle ,\left\langle W(\gamma
_{2})\right\rangle \}= \frac{4\pi }{\overline{k}}\left\langle W(\gamma
_{1}\circ \gamma _{2})(\hat{s})\right\rangle .
\end{equation}

\item For $SL(n,%
\mathbb{R}
)$ and $SU(n):$%
\begin{equation}
C_{\mathbf{12}}=2P_{\mathbf{12}}-\frac{2}{n}I_{\mathbf{12}}
\end{equation}%
and%
\begin{equation}
\begin{aligned}
\{\left\langle W(\gamma _{1};z)\right\rangle ,\left\langle W(\gamma
_{2};w)\right\rangle \} =&-4s(z,w)[\left\langle W(\gamma _{1}(\hat{s}%
);z)W(\gamma _{2}(\hat{s});w)\right\rangle  \\
&-\frac{1}{n}\left\langle W(\gamma _{1};z)\right\rangle \left\langle
W(\gamma _{2};w)\right\rangle ]. \label{M}
\end{aligned}
\end{equation}%
At the poles $z_{\pm }$,
\begin{equation}
\{\left\langle W(\gamma _{1})\right\rangle ,\left\langle W(\gamma
_{2})\right\rangle \}= \frac{4\pi }{\overline{k}}[\left\langle W(\gamma
_{1}\circ \gamma _{2})(\hat{s})\right\rangle -\frac{1}{n}\left\langle
W(\gamma _{1})\right\rangle \left\langle W(\gamma _{2})\right\rangle ].
\end{equation}

\item For $Sp(2n,%
\mathbb{R}
)$ and $SO(n):$%
\begin{equation}
C_{\mathbf{12}}=P_{\mathbf{12}}+\chi _{\mathbf{12}}
\end{equation}%
and%
\begin{equation}
\begin{aligned}
\{\left\langle W(\gamma _{1};z)\right\rangle ,\left\langle W(\gamma
_{2};w)\right\rangle \} =&-2s(z,w)[\left\langle W(\gamma _{1}(\hat{s}%
);z)W(\gamma _{2}(\hat{s});w)\right\rangle   \label{B} \\
&-\left\langle W(\gamma _{1}(\hat{s});z)W(\gamma _{2}^{-1}(\hat{s}%
);w)\right\rangle ].
\end{aligned}
\end{equation}%
At the poles $z_{\pm }$,
\begin{equation}
\{\left\langle W(\gamma _{1})\right\rangle ,\left\langle W(\gamma
_{2})\right\rangle \}= \frac{2\pi }{\overline{k}}[\left\langle W(\gamma
_{1}\circ \gamma _{2})(\hat{s})\right\rangle -\left\langle W(\gamma
_{1}\circ \gamma _{2}^{-1})(\hat{s})\right\rangle ].
\end{equation}
\end{itemize}
Above, $(\gamma _{1}\circ \gamma _{2})(\hat{s})$ denotes the composed
loop starting at the base (intersection) point $x^{i}(\hat{s})$, then
traveling along $\gamma _{1}$ until it reaches the beginning of the second
loop at $y^{i}(\hat{s})$, then following along $\gamma _{2}$ until it
reaches the point $y^{i}(\hat{s})=x^{i}(\hat{s})$, defining a single
curved with no base point. The curve $(\gamma _{1}\circ \gamma _{2}^{-1})(\hat{s})$
simply means that at the point $y^{i}(\hat{s})$ the second loop is
traveled in the reverse direction. See fig 1. in \cite{Hassan} for further reference. This is how the intersection point is resolved in terms of the
product of (rerouted) loops. In \eqref{A} and \eqref{B}, the composition of paths is done
using transport matrices with a different spectral parameter dependence in
their arguments. How to define a single Wilson loop in this situations is
not clear yet but the very form of the expression \eqref{z-Goldman} suggest itself it makes more sense as a quadratic algebra.

In principle, the algebra \eqref{pre-monodromy} should be identified as the precursor of the would-be Poisson algebra\footnote{There is a prescription for the Poisson algebra of the monodromy matrices of non-ultralocal integrable field theories originally proposed in \cite{Maillet} and applied to the lambda deformed PCM in \cite{YB-lambda}. However, this Poisson algebra, although useful for studying the involution properties of conserved charges extracted from the monodromy matrix, has three main drawbacks: i) it cannot be deduced directly from the Maillet algebra, ii) it only satisfy the Jacobi identity in a nested sense, see \cite{Maillet} and iii) its lattice version is not known. These three being the main issues associated to the non-ultralocality.} of the lambda model monodromy matrix $m(z)$ in \eqref{monodromy lambda} prior to the symplectic reduction. Schematically, we have the following relation among the $z$-dependent Poisson algebras considered so far:
\begin{equation}
\begin{array}{ccc}
\{A_{i}(x;z)_{\mathbf{1}},A_{j}(y;w)_{\mathbf{2}}\}_{\eqref{pre Maillet}} & \xrightarrow{\text{Sympl. Red.}} & \{A_{\sigma }(\sigma ;z)_{\mathbf{1}},%
A_{\sigma }(\sigma ^{\prime };w)_{\mathbf{2}}\}_{\eqref{Maillet}} \\ 
\bigg\downarrow \text{Pexp } &  & \bigg\downarrow \text{Pexp } \\ 
\{W(\gamma_{1} ;z)_{\mathbf{1}},W(\gamma_{2} ;w)_{\mathbf{2}}\}_{\eqref{pre-monodromy}} & \xrightarrow{\text{Sympl. Red.}}& \{m(z)_{\mathbf{1}},m(w)_{\mathbf{2}}\}=\text{?}. \label{alg square}
\end{array} 
\end{equation}

It is important to notice that the Wilson loop \eqref{Wilson z} is never trivial for generic values of $z$ even if we restrict it to the subset of
flat connections. Indeed, if we use the flatness conditions at $z_{\pm }$%
\begin{equation}
\partial _{i}A_{j(\pm )}-\partial _{j}A_{i(\pm )}+[A_{i(\pm )},A_{j(\pm )}]=0
\end{equation}%
simultaneously to calculate the 2-form $F(z)=\frac{1}{2}F_{ij}(z)dx^{i}\wedge dx^{j}$ required in the non-Abelian version of the Stokes theorem, we find that
\begin{equation}
F_{ij}(z)=\partial _{i}A_{j}(z)-\partial _{j}A_{i}(z)+[A_{i}(z),A_{j}(z)],
\end{equation}%
develops zeroes only at the poles $z_{\pm}$ with a general expression of the form
\begin{equation}
F_{ij}(z)=\varphi _{\lambda }^{-1}(z)X_{ij}(z).
\end{equation}%
Above, the $X_{ij}(z)$ denotes a combination of commutators of the components of $%
A_{i(\pm )}$ that never vanishes as can be checked explicitly. Then, the $z$-dependent Wilson loop \eqref{Wilson z} always depends on the area enclosed by $\gamma$ even on a surface with a trivial fundamental group like $D$. Clearly, at the points $z_{\pm }$, it is independent of $\gamma$ for flat connections $A_{(\pm)}$ and we get the trivial result. Let us notice though, that there are two ways to keep Wilson loops non-trivial on the disc: i) by introducing punctures as is usual in CS theory or ii) by introducing an spectral parameter dependence as we just did, where the twisting function now plays the r\^ole of an obstruction.

We now turn to the study gauge symmetries in the presence of $z$. Let us write the gauge moment \eqref{gauge Hamiltonian} in the form%
\begin{equation}
H(\epsilon )=\int\nolimits_{D}\left\langle \epsilon f\right\rangle
+\int\nolimits_{\partial D}\left\langle \epsilon \varphi \right\rangle ,
\end{equation}%
where%
\begin{equation}
f=\frac{k}{2\pi }(\Omega ^{T}F_{(+)}-F_{(-)}),\text{ \ \ }\varphi =-\frac{k}{%
2\pi }(\Omega ^{T}A_{(+)}-A_{(-)})
\end{equation}%
and%
\begin{equation}
\eta _{+}=\Omega \epsilon ,\text{ \ \ }\eta _{-}=\epsilon ,\text{ \ \ }%
\epsilon =\epsilon ^{(0)}+\epsilon ^{(1)}+\epsilon ^{(2)}+\epsilon ^{(3)}.
\end{equation}%
Similar as done in \eqref{Pos act}, we get now the action of gauge transformations on the 
$z$-dependent field \eqref{interpolating current}. Namely,%
\begin{equation}
\{A_{i}(x;z),H(\epsilon )\}=-\partial _{i}\epsilon
(x;z)-[A_{i}(x;z),\epsilon (x;z)]+2\varphi _{\lambda }^{-1}(z)X_{i}(x;z),
\end{equation}%
where we have defined $\epsilon (x;z)=\Omega (z/z_{-})\epsilon (x)$ and 
\begin{equation}
X_{i}(x;z)=[z_{+}z\varphi _{i}^{(0)},\epsilon ^{(1)}]-[z_{-}z^{-1}\varphi
_{i}^{(1)}+\varphi _{i}^{(2)},\epsilon ^{(2)}]-[\varphi _{i}^{(1)},\epsilon
^{(3)}].
\end{equation}%
Here, the $\varphi _{i}^{\prime }s$ are not to be interpreted as constraints, obviously. The
situation in the lambda model and the CS theory is thus the same: ordinary gauge
transformations do not extend outside the poles $z_{\pm }$ in a natural way.

Let us now consider the dressing group. Following its definition
\cite{dressing}, a dressing transformation is a $z$-dependent gauge transformation
preserving the analytic structure and the gauge connection nature of the the
Lax pair $\mathscr{L}_{\mu }(z)$, $\mu =0,1$ of the integrable field theory at hand. In the present case, this
translates into the following conditions on the $z$-dependent interpolating
field $A_{i}(x;z)$ and the dressing parameters $\theta _{\pm }(x;z)$,%
\begin{equation}
A_{i}^{\prime }(z)=-\partial _{i}\theta _{\pm }(z)\theta _{\pm }(z)^{-1}+\theta _{\pm
}(z)A_{i}(z)\theta _{\pm }(z)^{-1},\text{ \ \ }\theta(z) =\theta _{-}(z)^{-1}\theta _{+}(z).
\end{equation}%
Both relations are consistent provided%
\begin{equation}
\partial _{i}\theta(z) \theta(z) ^{-1}=-A_{i}(z)+\theta(z) A_{i}(z)\theta(z) ^{-1},
\end{equation}%
which is solved by taking
\begin{equation}
A_{i}(z)=-\partial _{i}\Psi(z) \Psi(z) ^{-1},\text{ \ \ }\theta(z) =\Psi(z) g(z)\Psi(z) ^{-1},
\end{equation}%
for $g(z)$ independent of the disc coordinates $x^{i}$. From this follows the expected result\footnote{%
The $g^{-1}_{\pm}(z)$ are required in order to preserve the normalization condition of
the wave function $\Psi(z)$, e.g. $\Psi (0;z)=1$ in the present situation.} 
\begin{equation}
\Psi ^{g}(x;z)=\theta _{\pm }(x;z)\Psi (x;z)g_{\pm }^{-1}(z).
\end{equation}%
In other words, the field $A_{i}(x;z)$ must be pure gauge in order for the dressing group to be naturally defined in $D$, but nothing
guarantees this is always fulfilled in our current situation. However,
under the symplectic reduction only the component of the gauge field along the boundary of $D$ survives and when complemented with the time component, the pair $A_{\mu}(\sigma;z)$, $\mu =0,1$ do satisfy such a flatness condition, as a consequence, we are prompted to conclude that the infinite-dimensional group of dressing symmetries is a
purely boundary effect in the $z$-dependent CS theory. In this way, we
recover the lambda model dressing group action considered before in 
\eqref{2.4} directly from the CS theory by using the key relations \eqref{key rel 1} and \eqref{key rel 2}.

Finally, if we give up the condition of preserving the analytic structure of the $z$-dependent current \eqref{interpolating current}, we can define formal $z$-dependent gauge transformations
\begin{equation}
\begin{aligned}
A_{i}^{g}(z)&=-\partial _{i}g(z)g(z)^{-1}+g(z) A_{i}(z)g(z)^{-1}, \\
W^{g}(z)&= g(x_{0};z)W(z)g(x_{0};z)^{-1}, \label{last}
\end{aligned}
\end{equation}%
where we have used \eqref{Wilson z} with a base point at, say, $x_{0}$. Unfortunately, we have not succeeded in finding the $z$-dependent gauge moment generalization of \eqref{gauge Hamiltonian} that generates the infinitesimal gauge transformations \eqref{last} in Hamiltonian form.  

\section{Concluding remarks}\label{conclusions}

In this paper we have shown, by employing the Hamiltonian and symplectic formalisms and by carefully taking into account the spectral parameter dependence $z$, that the classical integrable structure of the $AdS_{5}\times S^{5}$ superstring lambda model phase space is embedded in the phase space of a higher dimensional gauge theory corresponding to a double Chern-Simons theory and that it reveals itself as a byproduct of the symplectic reduction process applied to this double CS theory. We have also generalized the results initially pointed out in \cite{lambdaCS} and valid on the constraint surface (i.e. weakly) to results that are now valid on the whole phase space (i.e. strongly), making the connection between both theories quite precise. By trading the lambda model by the CS theory, the problematic non-ultralocality characteristic of the lambda models can be bypassed or avoided for any value of $\lambda$ at the cost of augmenting the number of Hamiltonian constraints by two, namely, $F_{(\pm)}\approx 0$. Once the symplectic reduction has been done and the lambda model has been recovered through the equations \eqref{key rel 1} and \eqref{key rel 2}, the remaining Hamiltonian constraints $\varphi \approx 0$ are still to be considered, the final stage in identifying the physical (or transverse) degrees of freedom of the closed string follows from the implementation of the dressing gauge, which fixes just those conjugacy classes of the monodromy matrix (at $z_{\pm}$) corresponding to the action of the kappa and ordinary gauge symmetries of the lambda model. In the de-compactification limit of an infinitely long string, the lambda-deformed giant-magnon spectrum is recovered \cite{lambda background}. The advantage of doing this is that in CS theories the symplectic reduction can be performed at the quantum level before the non-ultralocality, i.e. the $\delta^{\prime}_{\sigma \sigma^{\prime}}$ term in the Kac-Moody algebra, manifests itself. The new feature is the presence of $z$, \cite{in progress}.

In all this story we have found an interesting connection with the Goldman bracket, which certainly deserves further study. For instance, as presented in \eqref{z-Goldman}, it is not clear yet how to define the product of curves at the intersection points. This can be inferred from the fact that the Wilson loops and transport matrices depend on different arguments, $z$ and $w$, as in the PCM lambda model examples \eqref{A}, \eqref{M} and \eqref{B}, so in the eventual situation where we write the tensor Casimir $C_{\mathbf{12}}$ in the expression \eqref{proj s} in a particular matrix representation, it is not clear yet how to compose transport matrices with different spectral parameter dependence into a single object (although this does not seems to be mandatory). This also obscures the properties of the Poisson bracket \eqref{z-Goldman} under $z$-dependent gauge transformations, which should, in principle, obey a special property at the intersection points. We expect to consider these issues elsewhere.

It would be interesting to consider the inclusion of the spectral parameter $z$ in the 2+1 dimensional CS theory, where the disc is replaced by an arbitrary genus $g$ Riemann surface $\Sigma_{g}$ with several circle boundaries and study the resulting boundary integrable system induced by the symplectic reduction to the space of flat connections on $\Sigma_{g}$. Notice that the time evolution of each circle boundary of the surface $\Sigma_{g}$ in the CS theory is topologically equivalent to the world-sheet swept by the closed string in the lambda model as it evolves in time so, in principle, we can replace the disc by any $\Sigma_{g}$ with a single boundary and obtain the same integrable field theory at the boundary. What kind on non-trivial new features will emerge by doing this or by considering several circle boundaries deserves further study as well.

\section*{Acknowledgements}

This work was supported by the S\~ao Paulo Research Foundation (FAPESP) under the research grant 2017/25361-7. \\
The author thanks the organizers of the workshop ``Exactly Solvable Quantum Chains" held at IIP, Brasil, for their kind invitation to present part of the results reported in this work. \\
The author would like thank the referees for valuable comments and suggestions.




\end{document}